# Liquid Heterostructures: Generation of Liquid-Liquid Interfaces in Free-Flowing Liquid Sheets

David J. Hoffman[1], Hans A. Bechtel[2], Diego A. Huyke[3], Juan G. Santiago[3], Daniel P. Deponte[1], Jake D. Koralek[1]*

[1]Linac Coherent Light Source, SLAC National Accelerator Laboratory
Menlo Park, CA 94720, USA.
* Email: koralek@slac.stanford.edu

[2]Advanced Light Source, Lawrence Berkeley National Laboratory
Berkeley, CA 94720, USA.

[3]Department of Mechanical Engineering, Stanford University
Stanford, CA 94305, USA

**Abstract:** Chemical reactions and biological processes are often governed by the structure and transport dynamics of the interface between two liquid phases. Despite their importance, our microscopic understanding of liquid-liquid interfaces has been severely hindered by difficulty in accessing the interface through the bulk liquid. Here we demonstrate a method for generating large-area liquid-liquid interfaces within free-flowing liquid sheets, which we call liquid heterostructures. These sheets can be made thin enough to transmit photons from across the spectrum, which also minimizes the amount of bulk liquid relative to the interface and makes them ideal targets for a wide range of spectroscopies and scattering experiments. The sheets are produced with a microfluidic nozzle that impinges two converging jets of one liquid onto two sides of a third jet of another liquid. The hydrodynamic forces provided by the colliding jets both produce a multilayered laminar liquid sheet with the central jet is flattened in the middle. Infrared microscopy, white light reflectivity, and imaging ellipsometry measurements demonstrate that the buried layer has a tunable thickness and displays well-defined liquid-liquid interfaces, and that the inner layer can be thinner than 100 nm.

**I. Introduction**

Liquid-liquid interfaces are of key importance to biology, chemistry, and engineering. These interfaces can be the sites of reactions,[1] adsorption,[2] self-assembly,[3] and chemical or electron transfer.[4] Despite its significance, the liquid-liquid interface and its vicinity remains difficult to study, as the interface contains far fewer molecules than the bulk in typical systems. This asymmetry makes isolating spectroscopic or scattering signals from the interface challenging, and has inspired many surface-specific or surface-enhanced spectroscopies.[5, 6] The liquid-liquid interface could be better examined by minimizing the amount of liquid not in contact with an interface–effectively minimizing the bulk signal–which could enable the investigation of interfaces with non-specialized techniques.

Towards the goal of generating efficient targets for liquid-liquid interface studies, this work demonstrates the generation of liquid heterostructures–liquid sheet jets that containing thin, discrete layers of different liquids. Microfluidic liquid jets, and in particular sheet jets,[7-12] have been rapidly gaining popularity for liquid-phase spectroscopic experiments. The liquid sheet jets are flat, vacuum-stable, fast-flowing, free-standing laminar liquid structures that can have sub-micron thicknesses. Additionally, the sheets are rapidly self-refreshing and avoid accumulated sample or container damage from high intensity or ionizing radiation. These traits have led to their adoption for extreme ultraviolet/soft X-ray spectroscopies,[7, 9] ultrafast electron diffraction,[13, 14] and a range of other use cases such as high harmonic generation and high-intensity laser targets.[15-17] Conventional microfluidics studies have also demonstrated that well-defined liquid-liquid interfaces can form between immiscible fluids in laminar flow,[18,19] suggesting that the liquid sheet jets provide a possible platform for supporting liquid-liquid interfaces.

The liquid heterostructures demonstrated here were produced by impinging a cylindrical liquid jet with two flanking cylindrical jets of a second, immiscible liquid. This flow occurs inside a three-channel microfluidic nozzle, with the three microfluidic channels intersecting at the nozzle exit. When liquid flows through the outer channels at sufficiently high flow rates, the collision of the jets generates a leaf-shaped liquid sheet bounded by a thicker, cylindrical rim in the streamwise-transverse plane orthogonal to the plane of the original colliding jets (illustrated in Fig. 1).[20-23] The inner jet of a second liquid is flattened into a thin layer by the same hydrodynamic forces that make the outer sheet, while also being completely enveloped in the outer sheet. This process is analogous to the generation of a gas-accelerated liquid sheet jet from a cylindrical jet using the same type of microfluidic nozzle,[10] except with a second liquid replacing the pressurized gas. The produced heterostructure dimensions were found to depend on flow rates and liquid properties, as well as the ordering of the liquids.

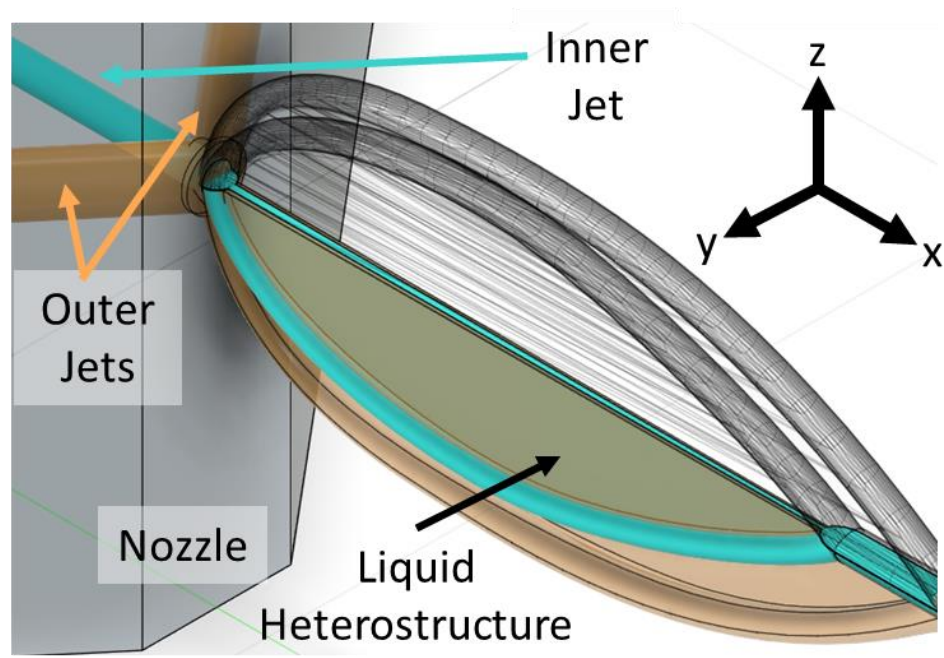

**Fig. 1.** Schematic cross-sectional diagram of the microfluidic nozzle and liquid heterostructure sheet jet. The three cylindrical jets (in the XY plane) meet at the end of the nozzle, where the hydrodynamic forces from the impinging outer jets (orange) generate a sheet in the XZ plane. The inner jet of an immiscible fluid (cyan) is flattened into a discrete layer within the sheet. The cyan rim of the sheet is thinly coated by the outer orange fluid, but is highlighted cyan for clarity.

The thicknesses of the liquid layers were measured with Fourier transform infrared (FTIR) microscopy, which provides spatially resolved absorption spectra that can be used to determine the total quantity of each liquid at each point in the sheet. The inner layer thickness was found to be tunable by adjusting relative flow rates, ranging from > 1 µm to < 100 nm, and followed similar scaling relations as the outer layers. The existence of distinct buried interfaces within the liquid sheet were then confirmed using reflectivity and imaging ellipsometry measurements. The buried interfaces provide additional surfaces for incident light to reflect from and generate thin-film interference effects. As the sheet layers vary in thickness with distance from the nozzle, the additional reflections manifest as intensity modulations in the sheet's thin film interference pattern. The reflection intensity modulations result in corresponding phase and polarization modulation in the ellipsometry measurements. These measurements were compared to the predicted ellipsometric observables and white-light reflection from an ideal heterostructure with layer thicknesses determined from FTIR measurements and showed excellent agreement. The results indicate the buried interfaces are sharp and smooth enough to produce coherent optical reflections–a high quality liquid heterostructure.

## II. Generating the Liquid Heterostructures

The liquid heterostructures were generated with the same borosilicate microfluidic chip nozzles that have been previously reported to generate ultrathin gas-accelerated liquid sheets[10] (lithographically manufactured by Micronit Microtechnologies BV). The microfluidic chip (shown schematically in Fig. 1) used for this study had outer channels (40 $\mu$m diameters) which meet an inner channel (20 $\mu$m diameter) at $\alpha \pm 40°$ angle The three channels meet at a distance of 25 $\mu$m from the exit of the nozzle. These convergent flows in the lithographically defined channels produce similar flows to colliding jets in air. Liquids were supplied to the microfluidic chip channels using two HPLC pumps (Shimadzu) connected to pulsation dampeners. The outer channels were supplied with one liquid at a combined flow rate, $Q_{out}$, ranging from 1.5 to 2.0 mL/min, while the inner channel was supplied a second liquid at a flow rate, $Q_{in}$, ranging up to 600 µL/min. For this work, the immiscible liquid pairs of water (Wat)/toluene (Tol) and water/cyclohexane (Cy) were chosen for the relatively high indices of refraction of toluene and cyclohexane[24] in visible wavelengths compared to water[25] (see Table I), which enables the reflectometry and ellipsometry experiments discussed in Section III.

The sheet morphology as a function of flow rate was monitored with microscopy. The white light reflection microscopy images (as seen in Fig. 2) were taken with a color CMOS camera (Motic, Moticam 2.0) through a 10× long working distance objective (NA 0.28) illuminated with white LEDs (Thorlabs LIUCWHA) at a 15° angle of incidence. For the reflection of white light from thin films, layers of different thicknesses will produce different colored bands due to the constructive or destructive interference at various wavelengths, and are discussed in detail in Section III.B.

In the absence of a liquid flowing through the center channel, this microfluidic chip forms a single-liquid sheet from the collision of two cylindrical jets (images in Supplemental Material). The liquid sheet thickness, $h$, created by the collision of two cylindrical jets has the following approximate form:[20-23]

$$h = \frac{aR^2}{r}. \tag{1}$$

Here, $R$ is the radius of the cylindrical jets, $r$ is the distance from the collision point, and $a$ is a scaling factor which depends on the specific geometry of the nozzle. Significantly, this thickness depends weakly on the velocity of the jets (or equivalently, the liquid flow rate) and is only dependent on the nozzle design. Increasing the liquid flow rate instead increases the total surface area of the liquid sheet, and can enable access to larger $r$ and thus smaller $h$. Similarly, it was found that liquids with lower air-liquid interfacial tensions (cyclohexane < toluene < water;[26, 27] see Table I) produced larger area sheets with access to smaller $h$ for the same flow rate, but the scaling factor *a* appeared to be unchanged.

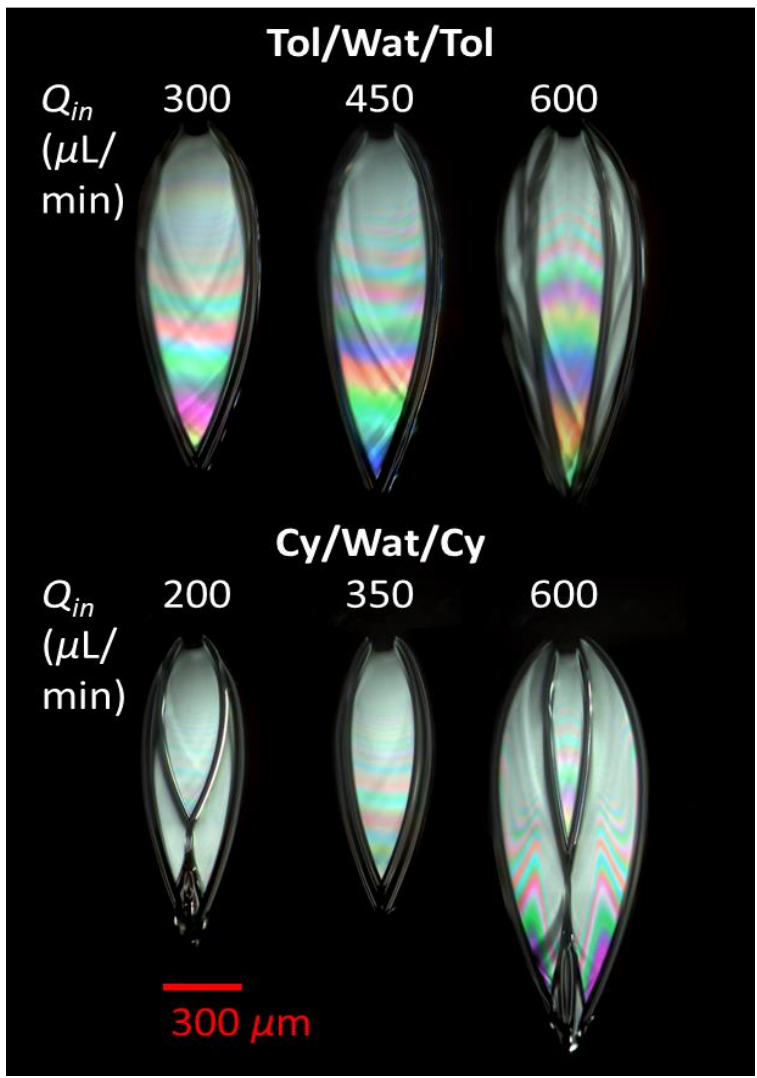

**Fig. 2.** White-light images of Tol/Wat/Tol (top) and Cy/Wat/Cy (bottom) liquid heterostructure sheets with outer flow rate, $Q_{out}$ = 1750 $\mu$L/min and a range of inner flow rates, $Q_{in}$. The sheets display a range of morphologies and thin film interference patterns depending on flow rate and fluid properties, but are all topologically identical to the schematic diagram shown in Fig. 1. Additional images can be seen in the Supplemental Material.

To produce the liquid heterostructure, a second, immiscible fluid was introduced through the inner channel. Representative Tol/Wat/Tol and Cy/Wat/Cy heterostructures are shown in Fig. 2 with fixed outer fluid flow rate, and inner fluid flow rate increasing to the right. At low inner-fluid flow rates (below the dripping/jetting transition, $Q_{in}$ ~150 $\mu$L/min), the sheet first exhibits instabilities and unsteady fluctuations, and at slightly higher flow rates becomes a stable laminar structure. Just prior to the fully stable sheet, half-formed structures are produced in which the inner sheet appears to breakup midway through the outer sheet. These structures are discussed in the supplement (S3) and may be analogous to the disintegrating/atomizing sheets observed for single fluid systems[28], but in this case the inner sheet is breaking up while still fully contained within the intact outer sheet. As the following sections will demonstrate, the fully formed stable sheets are ideal liquid heterostructures, which have a thin layer of the immiscible fluid sandwiched between two neighboring sheets.

Two visually distinct morphologies of immiscible liquid heterostructures can be identified. For structures where the interfacial tension between the inner and outer fluids is smaller than the interfacial tension between the outer fluid and air, we find that the rims of the two sheets appear to have similar shape and overlap. That is, the multi-liquid sheet will appear as a single sheet structure composed of fully overlapping layers, as can be seen for most of the Tol/Wat/Tol structures (Fig. 2, top row), or for sheets generated with water as the outer liquid (see Supplemental Material). On the other hand, when the interfacial tension between the two fluids significantly exceeds the interfacial tension of the outer fluid with air, then the rims of the inner sheet are not constrained by the rims of the naturally larger outer sheet, and will be laterally separated. In this morphology, the center sheet appears as a smaller sheet with rims fully contained by the larger outer sheet. For the immiscible pairs considered here, this latter case is most obvious with cyclohexane as the outer fluid and water as the inner fluid (Cy/Wat/Cy heterostructures, Fig. 2 bottom), as the interfacial tension of water with cyclohexane (~50 mN/m)[29] is twice the interfacial tension of cyclohexane with air (~25 mN/m).[27] In some cases, multiple leaf-shaped sheet structures are apparent for the center stream, with a second, shorter sheet structure (orthogonal to the first) and then a third sheet (parallel to the first) visible at the highest water flow rate in the Cy/Wat/Cy sheets in Fig. 2. Perhaps the most striking result of this work is that in both of these morphologies, the topology of the liquid heterostructure is the same, with the inner fluid always fully contained by the outer fluid even in cases where multiple inner sheets are formed. As will be examined with FTIR microscopy in the following section, the outer fluid is always observed to coat the entire inner structure, with fully laterally phase-separated structures never observed.

We note that it is also possible to use pairs of miscible liquids with this nozzle geometry. Due to the small length scales involved in the liquid sheets, this configuration provides for fast diffusive mixing between two liquids or solutions and allow for the monitoring of, e.g., fast chemical dynamics.[30] Similarly, phase transfer of solutes could be examined across immiscible interfaces. A full description of this capability will be discussed in a future publication.

## III. Sheet Measurement Results and Discussion

### A. FTIR Spectromicroscopy

Absorption measurements provide a direct measurement of the amount of material in a given optical path length, which can then be used to determine the thickness of layers in a liquid heterostructure. This procedure is a straightforward application of Beer's law:

$$A(\omega) = \varepsilon(\omega) c l \qquad (2)$$

where $A$ is the absorption at frequency $\omega$, $\varepsilon$ is the molar absorptivity of the relevant species at that frequency, $c$ is the concentration of the species, and $l$ is the optical path. As the liquid combinations used are essentially insoluble (with saturated solutions at room temperature of $< 0.1\%$ w/w for all cases) the concentrations are effectively the same as for neat solvents (55.5 M for water, 9.4 M for toluene, and 9.3 M for cyclohexane). With knowledge of the relevant molar absorptivities, the layer thicknesses of a given heterostructure can be readily determined from the measured absorbances. For this work, the integrated intensities of the absorbance peaks were considered to improve the signal-to-noise ratio.

While the common solvents discussed in this work (water, toluene, and cyclohexane) are negligibly absorbing in the visible spectrum, they have much stronger, spectrally well-separated vibrational transitions in the infrared (IR).[31, 32] The liquid heterostructures were imaged at beamline 2.4 at the Advanced Light Source (ALS), using an Agilent Cary 620 FTIR microscope fitted with a 15× all-reflective objective (NA= 0.62) and a 128×128 MCT pixel-array detector. This system simultaneously captures Fourier transform infrared (FTIR) spectra on each detector pixel, giving a spatial map of the IR spectra across the entire sheet. For these measurements, a globar source and KBr beamsplitter were employed. The FTIR spectrometer and microscope were purged with nitrogen, except for the space between the objectives. The liquid sheets were run in air in a custom sample cell with 1 mm thick $CaF_2$ windows connected to a ¼" vacuum drain hose. For some fluids and flow rates, the sheets were found to be thick enough at points to produce noticeable thin-film interference at infrared wavelengths (2.5 – 10 µm), which produced an oscillatory spectral background. This background was modeled and subtracted using the transfer matrix method (see Supplemental Material).[33, 34]

FTIR microscopy was used to study the composition of the complex morphologies that occur when the interfacial tension between the inner and outer fluids exceeds the interfacial tension between the outer fluid and air. In these cases, the inner layer in general cannot fill the full surface area of the sheet. Looking at the sheet formed by a Cy/Wat/Cy heterostructure, the spatial distribution of the component liquids was determined from the intensity of their characteristic vibrational modes. Fig. 3A shows a false-color image of the heterostructure with red indicating the intensity of a cyclohexane HCH bending mode[35] (~1450 cm$^{-1}$) and cyan indicating the intensity of the water OH stretch[36] (~3400 cm$^{-1}$). Spectra at two points inside and outside the inner layer sheet are shown in Fig. 3B. Water is confined entirely to the smaller-area layer, while cyclohexane is present throughout the entire structure. A similar result was found for all analogous morphologies measured for both Cy/Wat/Cy and Tol/Wat/Tol heterostructures (see Supplemental Material). This measurement provides evidence that the inner layer sheet is entirely surrounded by the outer layer as opposed to the sheet being laterally phase-separated.

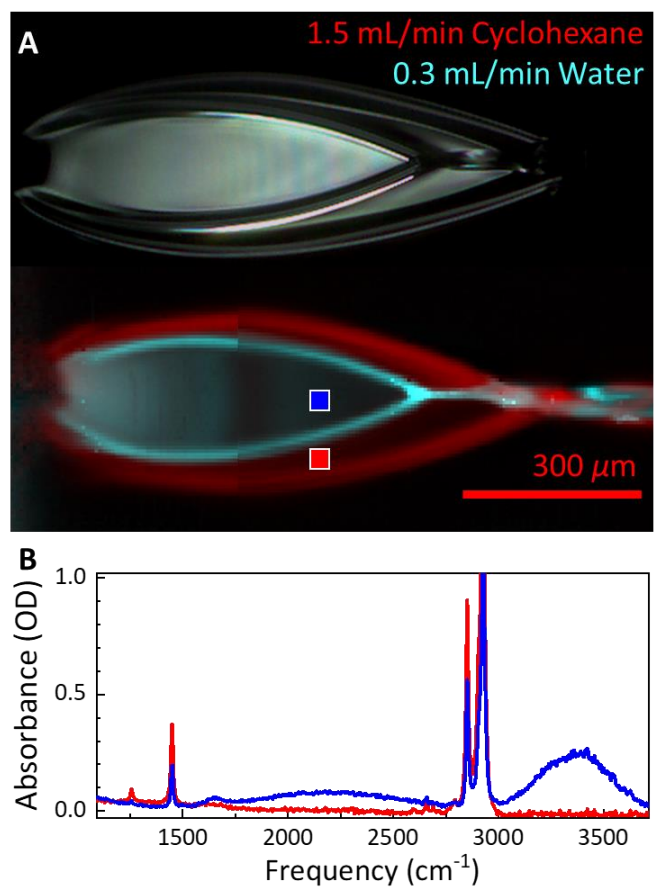

**Fig. 3. A.** White light image (top) and false-color IR microscopy image (bottom) of a Cy/Wat/Cy heterostructure ($Q_{out}$ = 1500 µL/min; $Q_{in}$ = 300 µL/min). Red indicates the intensity of a cyclohexane (1450 cm$^{-1}$ IR mode) and blue indicates the intensity of a water (3400 cm$^{-1}$ IR mode) in the false color image. Cyclohexane is found throughout the sheet but water is localized to the smaller interior sheet. **B.** Sample FTIR spectra at the points indicated in A. The sharp cyclohexane peaks are apparent in both, but the broad water peak at 3400 cm$^{-1}$ is only visible in the interior sheet.

The thickness of the layers, $h$, throughout the sheet as a function of flow rates was also examined. As the sheet flows away from the nozzle, we expect the sheet to get thinner according to the scaling law given in Eq. 1, resulting in smaller absorption peaks further from the nozzle (e.g., Fig. 4A). The measured absorption intensities could then be converted to thicknesses using Eq. 2. Sample thickness curves for different inner fluid flow rates, $Q_{in}$, with constant outer fluid flow rates, $Q_{out}$, are shown in Fig. 4B for Tol/Wat/Tol sheets. The oscillatory behavior in the data arises from imperfect subtraction of the thin-film interference background (see Supplemental Material). The absorption measurements show that the inner layer (blue, green, and violet points) also gets progressively thinner with distance from the nozzle, and that the layer thickness at a given point increases with increasing $Q_{in}$. Furthest from the nozzle and at low $Q_{in}$, the inner fluid layer thickness was found to be under 100 nm, with the smallest measured value of just over 30 nm. This thickness is comparable to those found for the ultrathin gas-accelerated sheets produced by the same nozzle.[10] The thickness of the outer layer (red, yellow, and brown points) was found to fit reasonably well to the theoretical model for a colliding sheet (eq. 1, red line in Fig. 4B), with a value of $a_{out} = 2.9 \pm 0.1$, and was found to be identical to the single-fluid colliding sheet value for the nozzle and essentially independent of both $Q_{in}$ and $Q_{out}$, as is typical for a colliding sheet.[20-23]

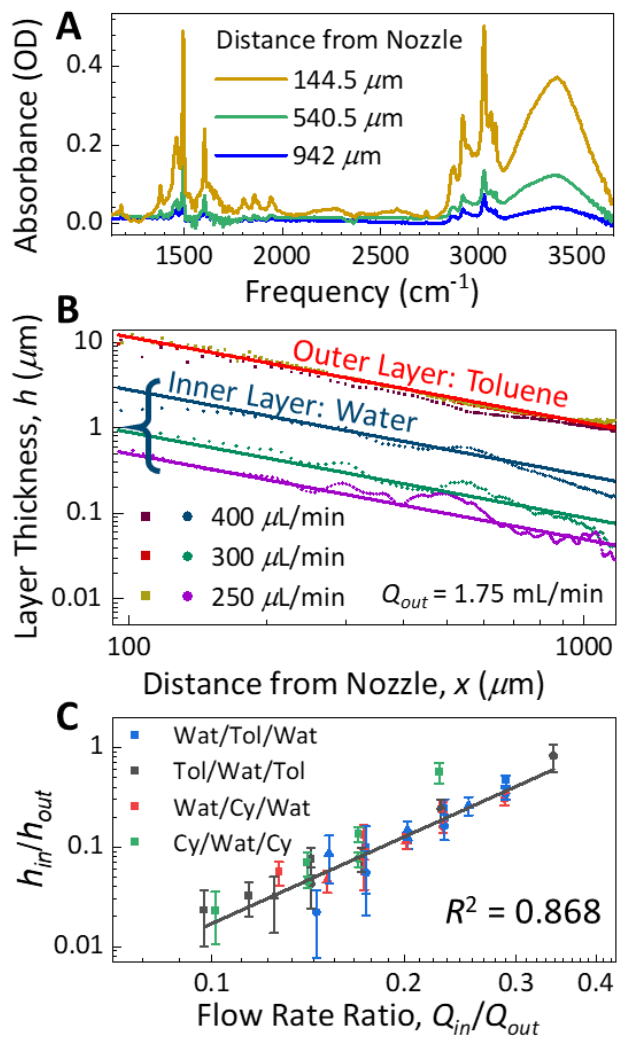

**Fig. 4. A.** Baseline-subtracted FTIR spectra of a Tol/Wat/Tol heterostructure ($Q_{out}$ = 1750 µL/min; $Q_{in}$ = 300 µL/min) at select distances from the nozzle. As the distance increases, the sheet layers become thinner, causing the toluene modes (sharp peaks) and water mode (broad peak at 3400 cm$^{-1}$) to decrease in intensity. **B.** Layer thicknesses in Tol/Wat/Tol heterostructures with $Q_{out}$ = 1750 µL/min and a range of $Q_{in}$. The outer layer thicknesses (red, yellow, and brown points) are independent of the flow rates, while the inner layer thicknesses increase with increasing $Q_{in}$ (blue, green, and violet points). All of the layers reasonably follow an $h \sim 1/x$ thickness dependence (Eq. 1, lines). **C.** Ratio of inner to outer layer thicknesses ($h_{in}/h_{out}$) plotted against the ratio of inner to outer fluid flow rates ($Q_{in}/Q_{out}$) for the studied heterostructures. Black line is a fitted power law with exponent $p = 2.94 \pm 0.17$.

Notably, the inner sheet appears to follow roughly the same scaling relationship as the outer layer, but with a different scaling factor. Sample lines of best fit to Eq. 1 are shown in Fig. 4B.

Unlike the largely invariant behavior of the outer layers, the corresponding values of $a_{in}$ for the inner layers have a dependence on both $Q_{in}$ and $Q_{out}$. In general, $a_{in}$ increases for increasing interior flow rate, $Q_{in}$, and decreases for increasing outer flow rate, $Q_{out}$. Plotting the ratio of the inner and outer fluid thicknesses ($h_{in}/h_{out}$) against the ratio of the inner and outer fluid flow rates ($Q_{in}/Q_{out}$) suggests conserved behavior across these regimes (Fig. 4C). Fitting this master curve to an empirical power law of the form

$$\frac{h_{in}}{h_{out}} = b \left( \frac{Q_{in}}{Q_{out}} \right)^p \tag{3}$$

gives values of $p = 2.94 \pm 0.17$ and $b = 14.4 \pm 3.9$ (black line in Fig. 4C). However, this relationship does not consider any fluid properties which affect the sheet morphologies and is likely incomplete.

**B. Ellipsometry and Reflectometry**

While it is clear from visible and IR microscopy that we can produce stable multi-liquid sheets, it is not immediately evident that these structures have well-defined buried liquid-liquid interfaces. Imaging ellipsometry and reflectometry were employed to interrogate these potential buried interfaces by looking for characteristic optical reflections from them. Generally speaking, if an interface is sharp and flat on the scale of the wavelength of incident photons it will be seen as an abrupt change in the index of refraction, which allows reflection to occur. If, on the other hand, the interface is smeared out or modulated, it will be seen as a smooth gradient in the index of refraction and suppress reflections like an anti-reflective coating.[37] If the experimental reflection intensities from a heterostructure match what is predicted for an ideal layered heterostructure (illustrated in Fig. 5A), then this provides strong evidence for well-defined buried liquid-liquid interfaces.

As was demonstrated with the IR measurements in the previous section, the thicknesses of the liquid layers in the heterostructure vary with distance from the nozzle. The changing thickness profile produces variation in the reflection intensity through thin film interference effects (illustrated in Fig. 5B). For the ideal heterostructure shown in Fig. 5A and 5B, there are three main limiting cases for the total reflection depending on the thicknesses of the component layers. First, all of the reflections from all four interfaces can fully destructively interfere to give a dark fringe. Second, the reflections from the two inner liquid-liquid interfaces can destructively interfere with each other, in which case the total reflection is solely made up of the reflections from the outer liquid-air interfaces. This case of no interference ("N") between the inner and outer interfaces is indicated with the dotted green line in Fig. 5B. The "N" line also corresponds to the reflection intensity for a single fluid sheet without an inner layer. Finally, the reflections from the inner fluid interfaces can constructively interfere with each other, and then alternately destructively or constructively interfere with the outer layer reflections ("D" and "C" regions respectively in Fig. 5B). In this case, adjacent bright fringes in the liquid heterostructure will show dramatically different intensities, as the changing layer thicknesses cycle the inner and outer reflections in and out of phase. The intensity maxima of alternating fringes can be used to generate two continuous curves that are 180° out of phase (Fig. 5B, red and blue dashed curves), which intersect at the no interference line. The fringes in the constructive and destructive interference regions have higher and lower reflectivities than the "N" line, respectively. The blue curve switches from constructive interference to destructive interference where it intersects the red curve at the "N" line, with the reverse happening for the red curve.

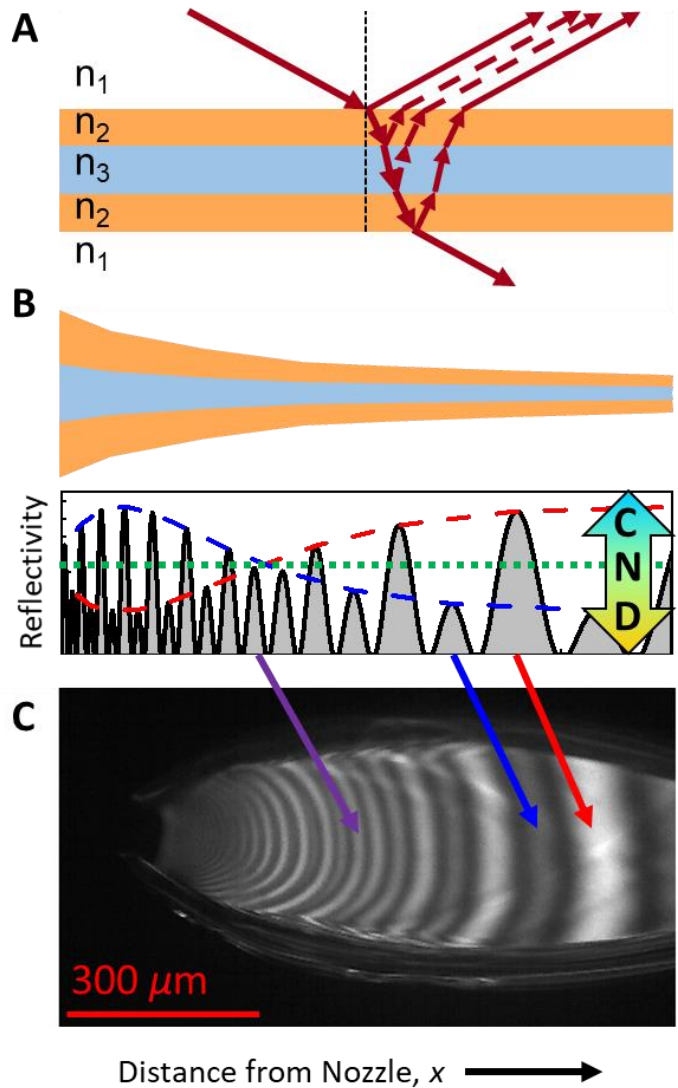

**Fig. 5. A.** Model of reflection off an ideal liquid heterostructure. The reflection intensity depends on thin film interference between the inner and outer layers. **B.** Model calculation of how the reflection of a monochromatic light source varies with distance from the nozzle ($x$) for a heterostructure with a $1/x$ thickness profile. Depending on the thickness of the layers, the reflections from the liquid-liquid interfaces can either not interfere (N), destructively interfere (D), or constructively interfere (C) with the liquid-air interface reflections. The peak intensities of alternating bright fringes can be used to construct two out-of-phase curves (red and blue dashed lines), which intersect the line of no interference (N, dotted green line) where they switch between constructive (C) and destructive (D). **C.** Reflection off a Tol/Wat/Tol heterostructure ($Q_{out}$ = 1750 $\mu$L/min; $Q_{in}$ = 300 $\mu$L/min) at a 30° angle of incidence using a 630 nm LED. The reflected fringe intensities show the interference effects illustrated in B, with the purple arrow indicating an N fringe and the red and blue arrows indicating representative C and D fringes, demonstrating reflections off buried liquid-liquid interfaces.

These characteristic thin film interference patterns can be seen clearly when a heterostructure is imaged using a monochromatic light source (630 nm LED bank, Thorlabs LIU630A). Fig. 5C demonstrates these patterns with an example Tol/Wat/Tol heterostructure ($Q_{out}$ = 1750 μL/min, $Q_{in}$ = 300 μL/min). Far from the nozzle, the bright fringes alternate between high and low intensity due to interference between the liquid-air and liquid-liquid interfaces (Fig. 5C, red and blue arrows). A similar effect is seen near the nozzle. The presence of these alternating fringes in the interference pattern requires the existence of buried liquid-liquid interfaces in the liquid heterostructure. Between these two regions, in the vicinity of the red and blue curve intersection point, neighboring bright fringes can be seen to have similar intensity (Fig. 5C, violet arrow), indicating a region where the inner liquid layer thickness results in destructive interference between the two liquid-liquid interface reflections (and thus no interference with the liquid-air interface reflections). While the thin film interference pattern indicates the presence of buried liquid-liquid interfaces, imaging ellipsometry was employed to quantitatively analyze the reflections from the liquid heterostructures.

Ellipsometry has been widely employed in the characterization of thin films.[38] The technique leverages differences in the reflectivity of different polarizations of light to gain information on the thickness, roughness, and dielectric properties of thin films. The ellipsometry technique measures the complex reflectance ratio $\rho = r_p/r_s$, where $r_p$ and $r_s$ are the complex-valued reflectivities of p-polarized (in plane of incidence) and s-polarized (perpendicular to plane of incidence) light. This quantity is typically expressed in terms of two ellipsometric angles:

$$\rho = \exp(i\Delta)\tan\Psi \tag{4}$$

where $\Delta$ is the phase difference between the reflected polarization components and $\Psi$ relates to the ratio of the *s* and *p* reflection amplitudes. Both measured ellipsometric angles are dependent upon the angle of incidence (AOI) of the light source. As $\rho$ is a ratio of reflectivities, the ellipsometric measurements do not require the careful intensity normalizations that are required for quantitative reflectometry measurements.

By combining ellipsometric methods with microscopy, it is possible to examine variations in the ellipsometric angles $\Delta$ and $\Psi$ across a surface.[39, 40] We employed an angle-resolved polarization-modulation imaging ellipsometer based on a literature design[40] which is detailed in the Supplemental Material. Briefly, the ellipsometer used the same CMOS camera and 10× long working distance objective (NA 0.28) used for the white light imaging shown in previous sections. A 630 nm LED bank passes through a linear polarizer set at 45°. The linearly polarized light reflects off the sheet with a different amplitude and phase for the reflected $s$ and $p$ components, depending on the angle of incidence and composition of the sheet. The reflection passes through the objective and through a quarter-wave plate on a motorized rotation-stage followed by a linear polarizer set at 90°. By collecting images of the sheet at five quarter-wave plate angles, it is possible to determine both ellipsometric angles at each point of the sheet in the focal plane (see Supplemental Material).

Using the imaging ellipsometer, the ellipsometric angles, $\Psi$ and $\Delta$, can be mapped across a liquid heterostructure at a range of AOIs. As was shown in Fig. 5C, the intensity of reflection from the heterostructure is modulated by interference between the liquid-liquid and air-liquid interfaces. This effect produces similar modulations in the ellipsometric observables. One phenomenon directly indicative of buried liquid-liquid interfaces can be seen in the phase difference, $\Delta$, between the $s$ and $p$ reflection components. A map of $\Delta$ across the example Tol/Wat/Tol heterostructure at a selection of AOIs can be seen in Fig. 6. In general, the liquid sheets act as near-ideal dielectric layers at 630 nm, where the total phase shift is 0° for the $s$ polarization and switches from 180° to 0° at Brewster's angle for the *p* polarization. However, the precise value of Brewster's angle for the heterostructure depends on the thickness of the inner liquid layer. At angles of incidence less than Brewster's angle for both component liquids, Δ is 180° across the sheet, and above Brewster's angle for both liquids Δ is 0° (Fig. 6, top and bottom respectively). However, between the two liquids' Brewster's angles, alternating fringes in the sheet will reflect with alternating phase (Fig. 6, middle). This phase modulation is an additional striking demonstration of the buried liquid-liquid interfaces in the liquid heterostructures.

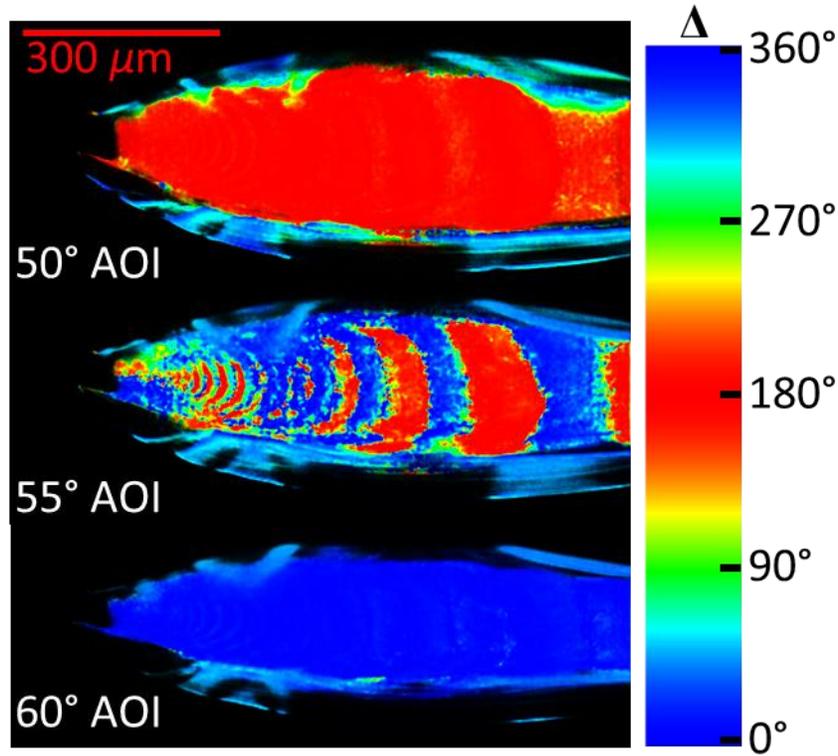

**Fig. 6**. Map of phase difference Δ between the *s* and *p* reflection components for different angles of incidence in a Tol/Wat/Tol heterostructure ($Q_{out}$ = 1750 µL/min; $Q_{in}$ = 300 µL/min). If the angle of incidence is less than (top image) or greater than (bottom image) Brewster's angle for both component liquids, Δ is uniformly 180° or 0°, respectively. Between the two component liquids' Brewster's angles (middle image), Δ can flip 180° between adjacent fringes due to interference effects.

A similar modulation is apparent in the measurement of Ψ at most AOIs. In general, the reflection modulation demonstrated in Fig. 5C will be different between the *s* and *p* polarization components, and result in a modulation of the reflected polarization angle (corresponding to the ellipsometric angle tan $\Psi = |r_p|/|r_s|$). A map of $\Psi$ across a liquid heterostructure (the Tol/Wat/Tol structure examined in Fig. 5C) at a 70° angle of incidence is shown in Fig. 7A, where cyan indicates a low value of $\Psi$ and yellow indicates a high value of $\Psi$. For most of the heterostructure, the angle of the reflected polarization alternates between high and low values, as neighboring bands produce alternating constructive or destructive interference between the outer and inner layers. This polarization modulation is once again clear evidence of sharp buried liquid-liquid interfaces within the sheet, which can be modeled to determine the layer dimensions.

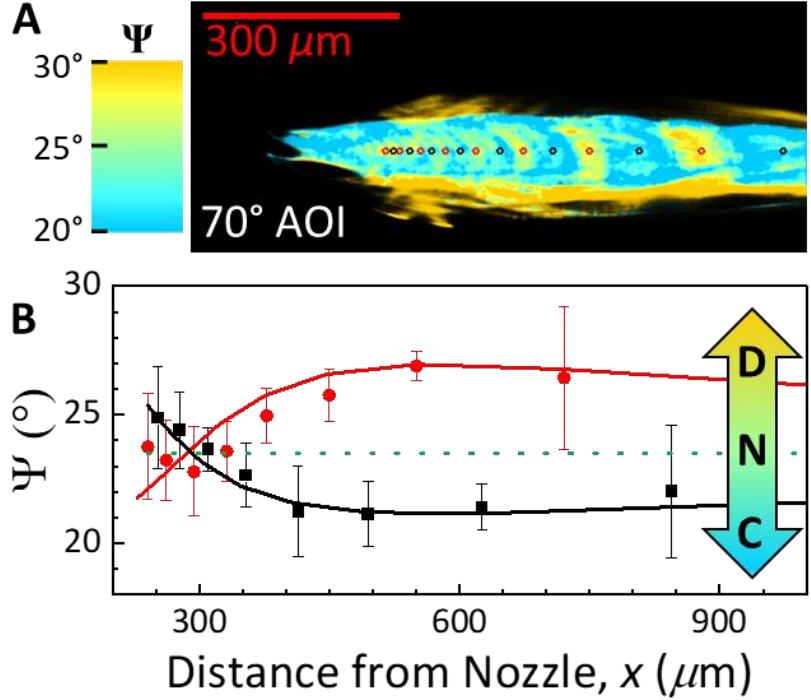

**Fig. 7. A.** Map of ellipsometric angle $\Psi$ (reflected polarization angle) for a Tol/Wat/Tol heterostructure ($Q_{out} = 1750$ µL/min; $Q_{in} = 300$ µL/min) at a 70° AOI. Constructive and destructive interference between reflections from the liquid/liquid and liquid/air interfaces result in alternating fringes of high and low $\Psi$. **B.** Comparison of measured and calculated values of $\Psi$ using the ideal heterostructure model (points and curves, respectively). Fringes are split into alternating pairs (red and black) corresponding to a 180° phase shift between the inner and outer reflections (analogous to the the red and blue curves in Fig. 5B). Calculated values derived from the liquid layer thicknesses measured by IR (Fig. 4B) using the ideal heterostructure model (Fig. 5A). Measured values are within error of the calculated values at all points. Dashed green line is the predicted value of $\Psi$ for a uniform toluene sheet.

**C. Combining Absorption and Reflection Measurements**

A full description of the liquid heterostructure can now be obtained by combining the reflection and absorbance measurements. Using the ellipsometry results in Fig. 7, layer thicknesses can be determined if a model structure is adopted (e.g., Fig. 5A). Complementarily, the absorption measurements from the FTIR microscopy provides information on the total amount of each liquid at every point in the sheet, but is agnostic to any layered structure. If the calculated layer thicknesses from the two methods correspond well with the amount of material in each point, this strongly supports the model heterostructure used for the reflection studies. This comparison can be most easily done by using the thicknesses determined by the IR measurements and calculating the ellipsometry observables from the ideal structure shown in Fig. 5A and 5B. The thicknesses of the layers were determined using the fits of the IR data to Eq. 1 (lines in Fig. 4B), with manual fine adjustment to the *a* parameters to better align to the observed locations of the bright fringes in the thin film interference pattern ($a_{out}$ = 2.99 and $a_{in}$ = 0.96, within error of the purely IR determined values of $a_{out}$ = 2.9 ± 0.1 and $a_{in}$ = 0.9 ± 0.2). For simplicity, it was assumed that the outer toluene layers were of equal thickness based on the symmetry of the nozzle. Ψ can then be calculated with the transfer matrix method[33, 34] (see Supplemental Material) for each point in the sheet using these layer thicknesses and the liquids' indices of refraction.

The measurements of Ψ are compared quantitatively with theory in Fig. 7B, where the points are the measured values at the peaks of the bright fringes, corresponding to the overlaid dots in Fig. 7A. Theoretical calculations for the model heterostructure of Fig 5B using the IR-determined thicknesses are shown as solid curves in Fig 7B. The changing layer thicknesses cycle the inner and outer reflections in and out of phase, which results in fringes with alternating values of Ψ. As was seen with the fringes' reflected intensities in Fig. 5B, these sets of alternating fringes can be used to construct two branches which are 180° out of phase with each other (Fig. 7B, red and black curves). As in Fig. 5B, the curves shown in Fig. 7B describe the calculated value of Ψ for fully constructive interference between the outer liquid-air interface reflections (i.e., interference fringe maxima, see Supplemental Material) which is then modulated by interference from the inner layer reflections. Analogous to the reflected intensities shown in Fig. 5B, the two curves intersect at the "N" line (Fig. 7B, dotted green line) at a distance of approx. 300 $\mu$m from the nozzle. At this point, the red curve transitions from describing constructive (< 300 $\mu$m, Ψ < 23.5°) to destructive (> 300 $\mu$m, Ψ > 23.5°) interference between the inner and outer layers, with the reverse occuring for the black curve. This branching behavior is a direct result of the buried interfaces, and is not seen for single-fluid sheets as indicated by the dashed line of constant Ψ in Fig. 7B.

These calculated values were compared to the measured values of Ψ around the most intense point of each bright fringe, which yield the highest signal to noise (Fig. 7A, red and black dots for the sets of alternating fringes). At all points, the measured values of Ψ for each fringe were within experimental error of the ideal heterostructure values (Fig. 7B), which corresponds to most of the sheet containing a water layer between 300 and 100 nm thick. Both sets of fringes also demonstrate the predicted pattern, where they start near the "No Interference" value close to the nozzle (predicted value for a single-fluid toluene sheet, dashed line in Fig. 7B) and show greater deviations at greater distance from the nozzle. The example Tol/Wat/Tol heterostructure is then quantitatively consistent with an ideal heterostructure with layer thicknesses corresponding to the values determined with the IR microscopy measurements. The experimentally determined cross-section for the example Tol/Wat/Tol heterostructure is shown in Fig. 8A, where the solid curves are the fit values used for the ellipsometry calculation shown in Fig. 7B and the points are the thicknesses determined from IR microscopy shown in Fig. 4. As a final validation of this determined structure, the colored fringes seen in the white light reflection images (e.g., Fig. 2) can be predicted from the profile shown in Fig. 8A using the transfer matrix method, the white light LED spectrum, and the wavelength-dependent indices of refraction (see Supplemental Material). While it is challenging to perfectly reproduce the camera's color and exposure settings, the calculated band colors provide a good quality match to the sheet's observed fringe colors (Fig. 8B). The combined white light reflection, ellipsometry measurements, and IR microscopy measurements then provide a consistent picture of a near-ideal liquid heterostructure containing optically flat and sharp liquid-liquid interfaces.

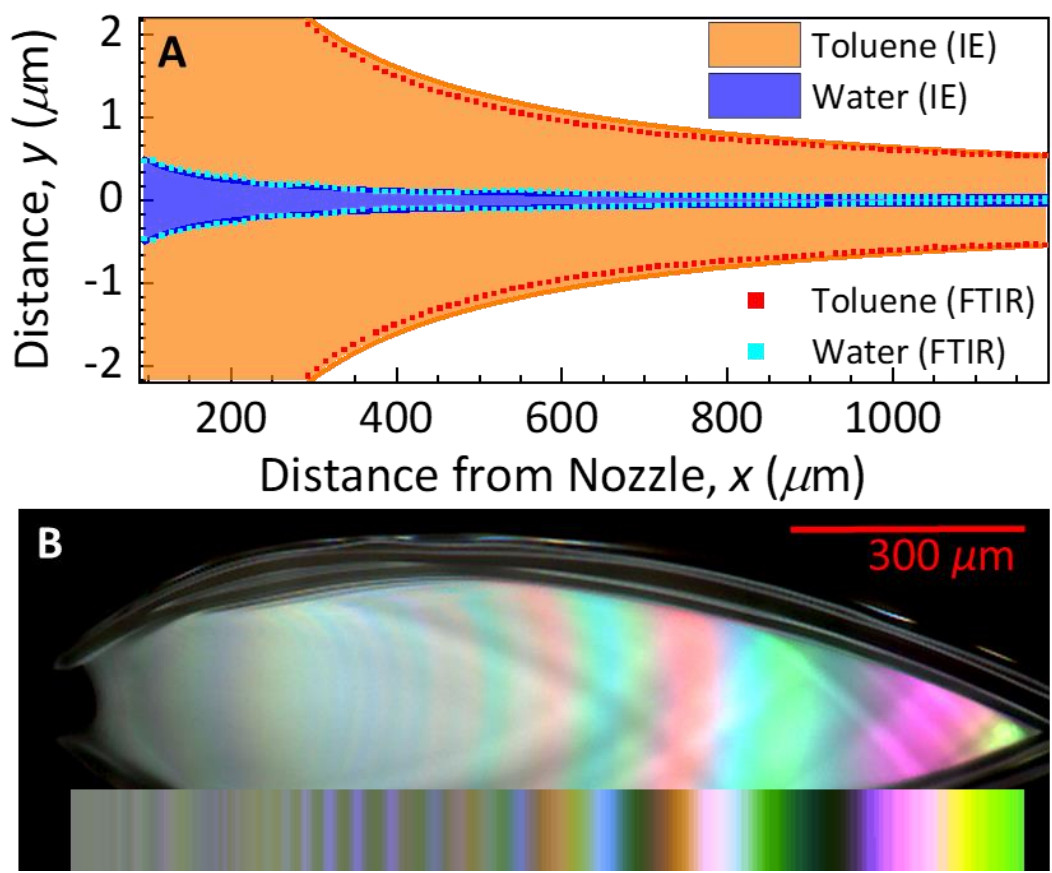

**Fig. 8. A.** Thickness profile for the Tol/Wat/Tol heterostructure ($Q_{out}$ = 1750 µL/min; $Q_{in}$ = 300 µL/min) from the ellipsometric model (solid curves) and FTIR absorbance (points). The two methods provide a common picture of the liquid heterostructure dimensions. **B.** White light image of the heterostructure compared to calculated white light reflection for an idealized heterostructure using the profile from A. The good agreement between colored bands provides additional validation for the cross-sectional profile.

## IV. Conclusion

This work demonstrated the generation of liquid heterostructures– free-flowing laminar layered liquid structures containing well-defined liquid-liquid interfaces – by impinging a cylindrical liquid jet with two jets of a second immiscible liquid using a microfluidic nozzle. Absorption measurements from FTIR microscopy demonstrated that the inner layer can be just 10s of nm thick and follows thickness scaling relationships similar to those seen for single-fluid colliding sheets and gas-accelerated sheets. Reflectivity and ellipsometry measurements demonstrated reflections off buried liquid-liquid interfaces, which indicate optically sharp and smooth interfaces. Excellent correspondence was found between the layer thicknesses determined by absorption measurements to those determined by reflectivity and ellipsometric measurements. These results indicate the liquid heterostructures formed are essentially ideal, and contain sharp, large-area, liquid-liquid interfaces.

As the inner layer can be tuned to be only 10s of nm thick, liquid heterostructures are ideal spectroscopic targets for interface studies, as the amount of bulk liquid not involved with the interface is greatly reduced. The free-flowing nature of the heterostructure also provides a number of unique benefits. Sheet jets have been used successfully for liquid studies in vacuum, which allows liquid-liquid interfaces to be studied with techniques that have been previously inaccessible due to the challenges of running liquids in vacuum. The liquid interface is also constantly refreshed, which allows for examination of kinetics of interface formation, as well as chemical and physical processes and reactions mediated by the interface on fast time scales (on the of order 100 $\mu$s, based on the jet velocity and sheet length). This unique collection of attributes makes the liquid heterostructure sheet jets a powerful new tool for examining the liquid-liquid interface.

**Acknowledgments**

This work was supported by the Department of Energy, Laboratory Directed Research and Development program at SLAC National Accelerator Laboratory, under contract DE-AC02-76SF00515. This research used resources of the Advanced Light Source, a U.S. DOE Office of Science User Facility under contract no. DE-AC02-05CH11231. D.A.H. is supported by a National Science Foundation Graduate Research Fellowship.

**Table I. Relevant Liquid Properties**

| Liquid | Interfacial Tension with Air (N/m) | Interfacial Tension with Water (N/m) | Index of Refraction at $\lambda = 630$ nm |
|---|---|---|---|
| **Water** | 71.9[26] | -- | 1.33[25] |
| **Toluene** | 28.5[27] | 35.8[29] | 1.49[24] |
| **Cyclohexane** | 25.2[27] | 50.0[29] | 1.43[24] |

**References**

1. Piradashvili, K.; Alexandrino, E. M.; Wurm, F. R.; Landfester, K., Reactions and Polymerizations at the Liquid–Liquid Interface. *Chem. Rev.* **2016,** *116*, 2141-2169.

2. Benjamin, I., Molecular Structure and Dynamics at Liquid-Liquid Interfaces. *Annu. Rev. Phys. Chem.* **1997,** *48*, 407-451.

3. Booth, S. G.; Dryfe, R. A. W., Assembly of Nanoscale Objects at the Liquid/Liquid Interface. *J. Phys. Chem. C* **2015,** *119*, 23295-23309.

4. Benjamin, l., Mechanism and Dynamics of Ion Transfer Across a Liquid-Liquid Interface. *Science* **1993,** *261*, 1558-1560.

5. Eisenthal, K. B., Equilibrium and Dynamic Processes at Interfaces by Second Harmonic and Sum Frequency Generation. *Annu. Rev. Phys. Chem.* **1992,** *43*, 627-661.

6. Perera, J. M.; Stevens, G. W., Spectroscopic studies of molecular interaction at the liquid–liquid interface. *Anal. Bioanal. Chem* **2009,** *395*, 1019-1032.

7. Ekimova, M.; Quevedo, W.; Faubel, M.; Wernet, P.; Nibbering, E. T. J., A liquid flatjet system for solution phase soft-x-ray spectroscopy. *Struct. Dyn.* **2015,** *2*, 054301.

8. Galinis, G.; Strucka, J.; Barnard, J. C. T.; Braun, A.; Smith, R. A.; Marangos, J. P., Micrometer-thickness liquid sheet jets flowing in vacuum. *Rev. Sci. Instrum.* **2017,** *88*, 083117.

9. Fondell, M.; Eckert, S.; Jay, R. M.; Weniger, C.; Quevedo, W.; Niskanen, J.; Kennedy, B.; Sorgenfrei, F.; Schick, D.; Giangrisostomi, E.; Ovsyannikov, R.; Adamczyk, K.; Huse, N.; Wernet, P.; Mitzner, R.; Föhlisch, A., Time-resolved soft X-ray absorption spectroscopy in transmission mode on liquids at MHz repetition rates. *Struct. Dyn.* **2017,** *4*, 054902.

10. Koralek, J. D.; Kim, J. B.; Brůža, P.; Curry, C. B.; Chen, Z.; Bechtel, H. A.; Cordones, A. A.; Sperling, P.; Toleikis, S.; Kern, J. F.; Moeller, S. P.; Glenzer, S. H.; DePonte, D. P., Generation and characterization of ultrathin free-flowing liquid sheets. *Nat. Commun.* **2018,** *9*, 1353.

11. Ha, B.; DePonte, D. P.; Santiago, J. G., Device design and flow scaling for liquid sheet jets. *Phys. Rev. Fluids* **2018,** *3*, 114202.

12. Menzi, S.; Knopp, G.; Al Haddad, A.; Augustin, S.; Borca, C.; Gashi, D.; Huthwelker, T.; James, D.; Jin, J.; Pamfilidis, G.; Schnorr, K.; Sun, Z.; Wetter, R.; Zhang, Q.; Cirelli, C., Generation and simple characterization of flat, liquid jets. *Rev. Sci. Instrum.* **2020,** *91*, 105109.

13. Nunes, J. P. F.; Ledbetter, K.; Lin, M.; Kozina, M.; DePonte, D. P.; Biasin, E.; Centurion, M.; Crissman, C. J.; Dunning, M.; Guillet, S.; Jobe, K.; Liu, Y.; Mo, M.; Shen, X.; Sublett, R.; Weathersby, S.; Yoneda, C.; Wolf, T. J. A.; Yang, J.; Cordones, A. A.; Wang, X. J., Liquid-phase mega-electron-volt ultrafast electron diffraction. *Struct. Dyn.* **2020,** *7*, 024301.

14. Yang, J.; Dettori, R.; Nunes, J. P. F.; List, N. H.; Biasin, E.; Centurion, M.; Chen, Z.; Cordones, A. A.; Deponte, D. P.; Heinz, T. F.; Kozina, M. E.; Ledbetter, K.; Lin, M.-F.; Lindenberg, A. M.; Mo, M.; Nilsson, A.; Shen, X.; Wolf, T. J. A.; Donadio, D.; Gaffney, K. J.; Martinez, T. J.; Wang, X., Direct observation of ultrafast hydrogen bond strengthening in liquid water. *Nature* **2021,** *596*, 531-535.

15. George, K. M.; Morrison, J. T.; Feister, S.; Ngirmang, G. K.; Smith, J. R.; Klim, A. J.; Snyder, J.; Austin, D.; Erbsen, W.; Frische, K. D.; Nees, J.; Orban, C.; Chowdhury, E. A.; Roquemore, W. M., High-repetition-rate (>kHz) targets and optics from liquid microjets for high-intensity laser–plasma interactions. *High Power Laser Sci. Eng.* **2019,** *7*, e50.

16. Yin, Z.; Luu, T. T.; Wörner, H. J., Few-cycle high-harmonic generation in liquids: in-operando thickness measurement of flat microjets. *J. Phys. Photonics* **2020,** *2*, 044007.

17. Obst, L.; Göde, S.; Rehwald, M.; Brack, F.-E.; Branco, J.; Bock, S.; Bussmann, M.; Cowan, T. E.; Curry, C. B.; Fiuza, F.; Gauthier, M.; Gebhardt, R.; Helbig, U.; Huebl, A.; Hübner, U.; Irman, A.; Kazak, L.; Kim, J. B.; Kluge, T.; Kraft, S.; Loeser, M.; Metzkes, J.; Mishra, R.; Rödel, C.; Schlenvoigt, H.-P.; Siebold, M.; Tiggesbäumker, J.; Wolter, S.; Ziegler, T.; Schramm, U.; Glenzer, S. H.; Zeil, K., Efficient laser-driven proton acceleration from cylindrical and planar cryogenic hydrogen jets. *Sci. Rep.* **2017,** *7*, 10248.

18. Atencia, J.; Beebe, D. J., Controlled microfluidic interfaces. *Nature* **2005,** *437*, 648-655.

19. Ismagilov, R. F.; Stroock, A. D.; Kenis, P. J. A.; Whitesides, G.; Stone, H. A., Experimental and theoretical scaling laws for transverse diffusive broadening in two-phase laminar flows in microchannels. *Appl. Phys. Lett.* **2000,** *76*, 2376-2378.

20. Taylor, G. I., Formation of thin flat sheets of water. *Proc. R. Soc. Lond. A* **1960,** *259*, 1-17.

21. Choo, Y. J.; Kang, B. S., Parametric study on impinging-jet liquid sheet thickness distribution using an interferometric method. *Exp. Fluids* **2001,** *31*, 56-62.

22. Bush, J. W. M.; Hasha, A. E., On the collision of laminar jets: fluid chains and fishbones. *J. Fluid Mech.* **2004,** *511*, 285-310.

23. Hasson, D.; Peck, R. E., Thickness distribution in a sheet formed by impinging jets. *AlChE J.* **1964,** *10*, 752-754.

24. Kozma, I. Z.; Krok, P.; Riedle, E., Direct measurement of the group-velocity mismatch and derivation of the refractive-index dispersion for a variety of solvents in the ultraviolet. *J. Opt. Soc. Am. B* **2005,** *22*, 1479.

25. Daimon, M.; Masumura, A., Measurement of the refractive index of distilled water from the near-infrared region to the ultraviolet region. *Appl. Opt.* **2007,** *46*, 3811.

26. Lange, N. A.; Dean, J. A., *Lange's Handbook of Chemistry*. McGraw-Hill: 1973.

27. Korosi, G.; Kovats, E. S., Density and surface tension of 83 organic liquids. *J. Chem. & Eng. Data* **1981,** *26*, 323-332.

28. Ryan, H. M.; Anderson, W. E.; Pal, S.; Santoro, R. J., Atomization characteristics of impinging liquid jets. *J. Propul. Power* **1995,** *11*, 135-145.

29. Backes, H. M.; Jing Jun, M.; E., B.; G., M., Interfacial tensions in binary and ternary liquid—liquid systems. *Chem. Eng. Sci.* **1990,** *45*, 275-286.

30. Schewe, H. C.; Credidio, B.; Ghrist, A. M.; Malerz, S.; Ozga, C.; Knie, A.; Haak, H.; Meijer, G.; Winter, B.; Osterwalder, A., Imaging of chemical kinetics at the water-water interface in a free-flowing liquid flat-jet. *arXiv preprint* **2022,** *arXiv:2201.05008*

31. Hale, G. M.; Querry, M. R., Optical Constants of Water in the 200-nm to 200-μm Wavelength Region. *Appl. Opt.* **1973,** *12*, 555.

32. Myers, T. L.; Tonkyn, R. G.; Danby, T. O.; Taubman, M. S.; Bernacki, B. E.; Birnbaum, J. C.; Sharpe, S. W.; Johnson, T. J., Accurate Measurement of the Optical Constants n and k for a Series of 57 Inorganic and Organic Liquids for Optical Modeling and Detection. *Appl. Spectrosc.* **2018,** *72*, 535-550.

33. Harbecke, B., Coherent and incoherent reflection and transmission of multilayer structures. *Appl. Phys. B* **1986,** *39*, 165-170.

34. Byrnes, S. J., Multilayer optical calculations. *arXiv preprint* **2016,** *arXiv:1603.02720* 1-20.

35. Takahashi, H.;  Shimanouchi, T.;  Fukushima, K.; Miyazawa, T., Infrared spectrum and normal vibrations of cyclohexane. *J. Mol. Spectrosc.* **1964,** *13*, 43-56.

36. Venyaminov, S. Y.; Prendergast, F. G., Water (H2O and D2O) Molar Absorptivity in the 1000–4000 cm−1Range and Quantitative Infrared Spectroscopy of Aqueous Solutions. *Anal. Biochem.* **1997,** *248*, 234-245.

37. Minot, M. J., The angluar reflectance of single-layer gradient refractive-index films. *J. Opt. Soc. Am.* **1977,** *67*, 1046.

38. Tompkins, H.; Irene, E. A., *Handbook of ellipsometry*. William Andrew: 2005.

39. Asinovski, L.;  Beaglehole, D.; Clarkson, M. T., Imaging ellipsometry: quantitative analysis. *Phys. Status Solidi A* **2008,** *205*, 764-771.

40. Chou, C.;  Teng, H.-K.;  Yu, C.-J.; Huang, H.-S., Polarization modulation imaging ellipsometry for thin film thickness measurement. *Opt. Commun.* **2007,** *273*, 74-83.

**Liquid Heterostructures: Generation of Liquid-Liquid Interfaces in Free-Flowing Liquid Sheets**

David J. Hoffman[1], Hans A. Bechtel[2], Diego A. Huyke[3], Juan G. Santiago[3], Daniel P. Deponte[1], Jake D. Koralek[1]*

[1]Linac Coherent Light Source, SLAC National Accelerator Laboratory
Menlo Park, CA 94720, USA.
* Email: koralek@slac.stanford.edu

[2]Advanced Light Source, Lawrence Berkeley National Laboratory
Berkeley, CA 94720, USA.

[3]Department of Mechanical Engineering, Stanford University
Stanford, CA 94305, USA

**Supplemental Material**

**S1. Liquid Sheet Surface Imperfections**

**S2. Water/Hydrocarbon/Water Heterostructures**

**S3. Stable-Unstable Transition in Heterostructures and Possible Emulsion Generation**

**S4. Additional FTIR Microscopy Images of Heterostructures**

**S5. Thin Film Calculations: Transfer Matrix Method**

**S6. Polarization-Modulation Imaging Ellipsometry Methods**

## S1. Liquid Sheet Surface Imperfections

Many of the liquid sheets produced by these microfluidic nozzles (in particular the single-liquid colliding sheets but also many of the heterostructures) have noticeable surface imperfections as a result of channel roughness and the nozzle exit geometry. An example of these effects are shown in Fig. S1. These surface imperfections can be seen to not dramatically affect the interference band structure in the sheet (meaning the size of the features has to be small in comparison the wavelengths of visible light, <100 nm), but are large enough to change the local curvature of the sheet. On illumination, these structures then create bright and dark areas in the sheet's reflection. These surface imperfections can similarly distort the ellipsometry measurements by creating reflections with different angles of incidence, which can limit the regions of the sheet which can be accurately measured.

These features do not contain noticeably more or less liquid than the rest of the sheet. This fact can be confirmed by looking at the FTIR microscopy of a single fluid sheet in comparison to the white light reflection of a sheet displaying significant surface imperfections (Fig. S1; single fluid sheet with water in all three channels: $Q_{out}$ = 1750 μL/min and $Q_{in}$ = 200 μL/min). The IR false-color image displays the intensity of the water HOH bend (1600 cm$^{-1}$)[1] across the sheet, corresponding to the sheet thickness. The distinct surface imperfections in the white light image do not register at all in the IR microscopy data, while the IR is able to detect a thickness variation near the bottom of the sheet which is obscured in the white light image by the surface imperfections. As the height of the surface imperfections are comparable to or smaller than the noise in the IR measurements, this constrains the height of these imperfections to ≤ 35 nm.

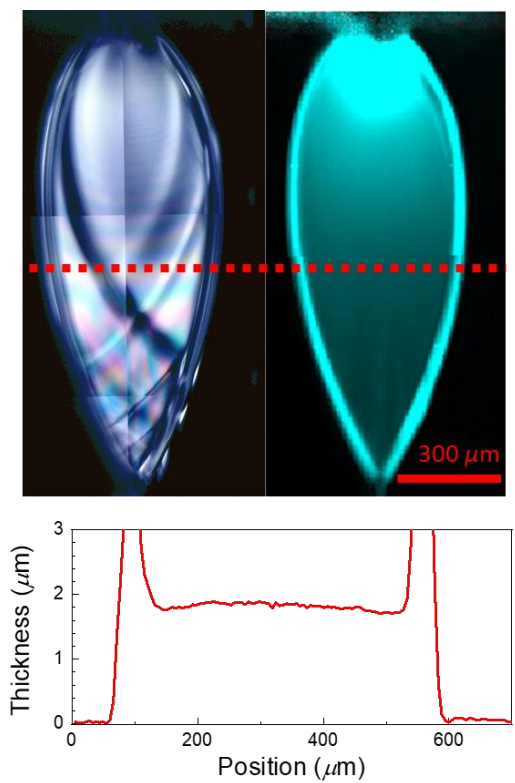

**Fig. S1.** White light (left) and IR false color image of a pure water sheet. The bright and dark surface imperfections in the white light image do not appear in the IR image, indicating that they correspond to only a small change in the thickness of the sheet (≤ 35 nm). Bottom: IR thickness data for sheet cross section indicated by dashed line.

## S2. Water/Hydrocarbon/Water Heterostructures

While the main text focused on water sheets within hydrocarbon sheets, the heterostructures could also be generated such that a thin layer of hydrocarbon was placed within a water sheet. Stable structures were generated with both an inner toluene layer and with an inner cyclohexane layer. Sample heterostructures with $Q_{out}$ = 1750 μL/min can be seen in Fig. S2. As the interfacial tension between water and air (72 N/m)[2] is much greater than the interfacial tension between water and toluene (36 N/m) or cyclohexane (50 N/m),[3] at all stable flow rates examined the sheets presented the simple morphology, wherein the rims of the sheet jets overlap and all of the liquid layers cover the same area. This is analogous to the Tol/Wat/Tol sheets with a low water flow rate presented in the main text. The water/hydrocarbon/water heterostructures

S.3

still had substantially smaller surface areas than the pure water colliding sheet with the same outer fluid flow rate (Fig. S2 top), with the cyclohexane-containing sheets slightly smaller than the toluene-containing sheets. The interfacial tension between the liquids likely plays a role in the area of the heterostructures in these cases, similar to how the liquid/air interfacial tension impacts the area of the single-liquid sheets.

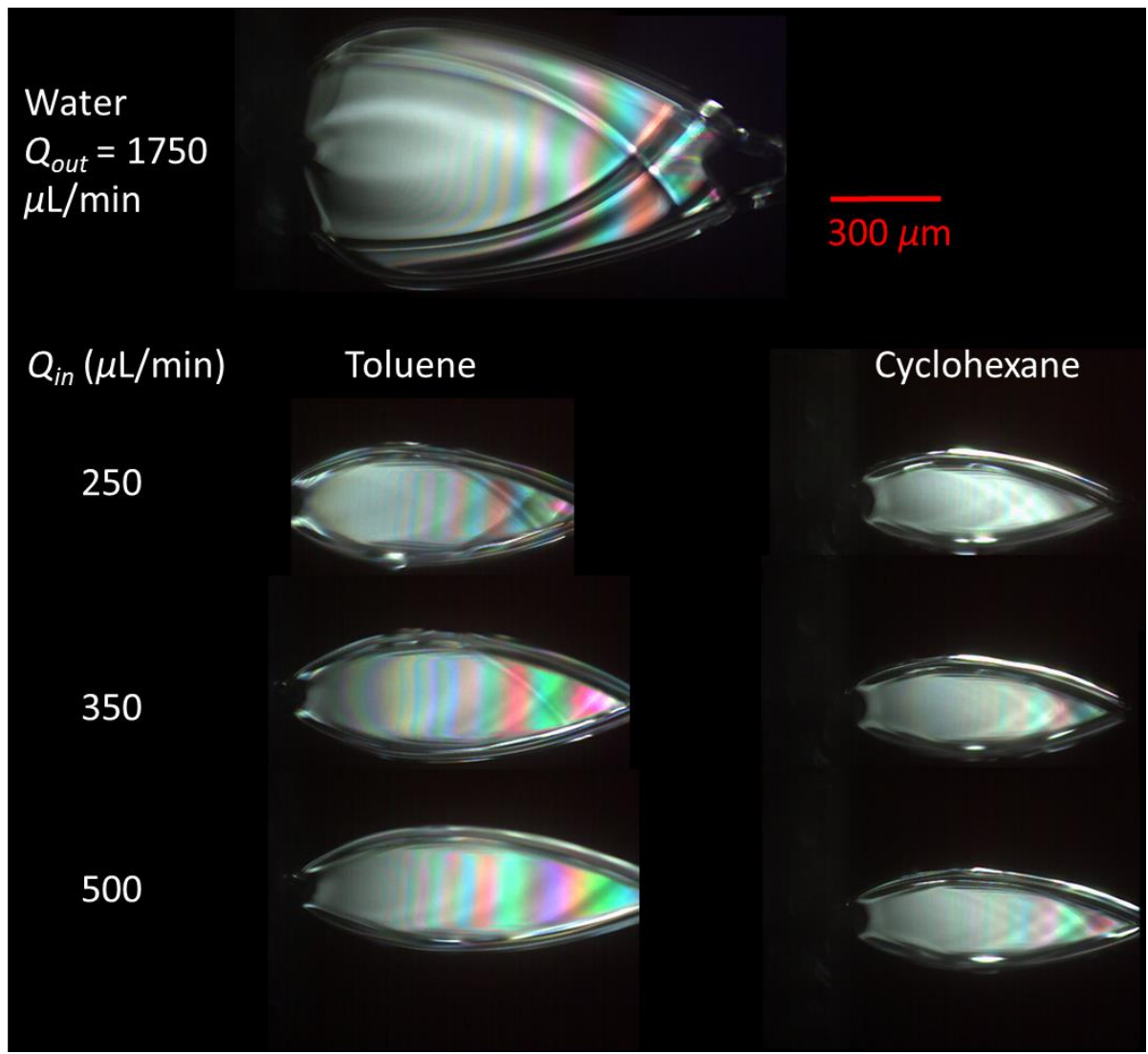

**Fig. S2.** White light images of a water colliding sheet (Top) and various Water/Hydrocarbon/Water heterostructures (bottom) with toluene and cyclohexane.

**S3. Stable-Unstable Transition in Heterostructures and Possible Emulsion Generation**

Below a critical flow rate of the inner fluid, the stable liquid heterostructure cannot form. There appear to be two main transitions before the fully stable liquid heterostructure can form. If $Q_{in}$ is below the flow rate which would produce a stable jet in the absence of the outer fluids (i.e., the nozzle would be periodically dripping, $Q_{in} < \sim 150$ μL/min for this nozzle), the resulting sheet was found to appear unstable, where the sheet would reflect diffuse white light instead of specular reflections typical of the stable sheets. As the inner flow rate increased, a more stable structure would form, but the sheets would also contain complicated spanwise wave patterns. These states are shown in Fig. S3 for Tol/Wat/Tol heterostructures, in the section labeled "Spanwise Waves."

Just above the flow rate necessary for a stable jet ($Q_{in} \sim 150$ μL/min), a well-formed sheet structure could be produced. However, at some distance from the nozzle, there would be a sudden shift from specular reflection to diffuse white reflection. The front of the stable-unstable transition in the sheet moved further from the nozzle as $Q_{in}$ increases (Fig. S3, "Atomizing"). As was briefly discussed in the main text, this appears to be analogous to the atomization process which occurs when a liquid sheet from impinging liquid jets disintegrates at high flow rates into aerosolized droplets.[4] Similar behavior was seen in the gas-accelerated liquid sheets at very high gas loads as well.[5] As the outer sheet structure is not completely disrupted, it appears that the inner liquid is still confined to the sheet structure. These structures may then allow for the generation of emulsions in liquid sheets.

The morphology of the heterostructure may significantly affect the resulting emulsion. For Cy/Wat/Cy sheets, the water rims are towards the center of the cyclohexane sheet, as opposed to being coincident with the outer fluid rims in the Tol/Wat/Tol sheets shown in Fig. S3.

The water rims in the Cy/Wat/Cy sheets then disintegrate in the center of the sheet (Fig. S4), which may provide a much higher hydrocarbon loading in the sheet center.

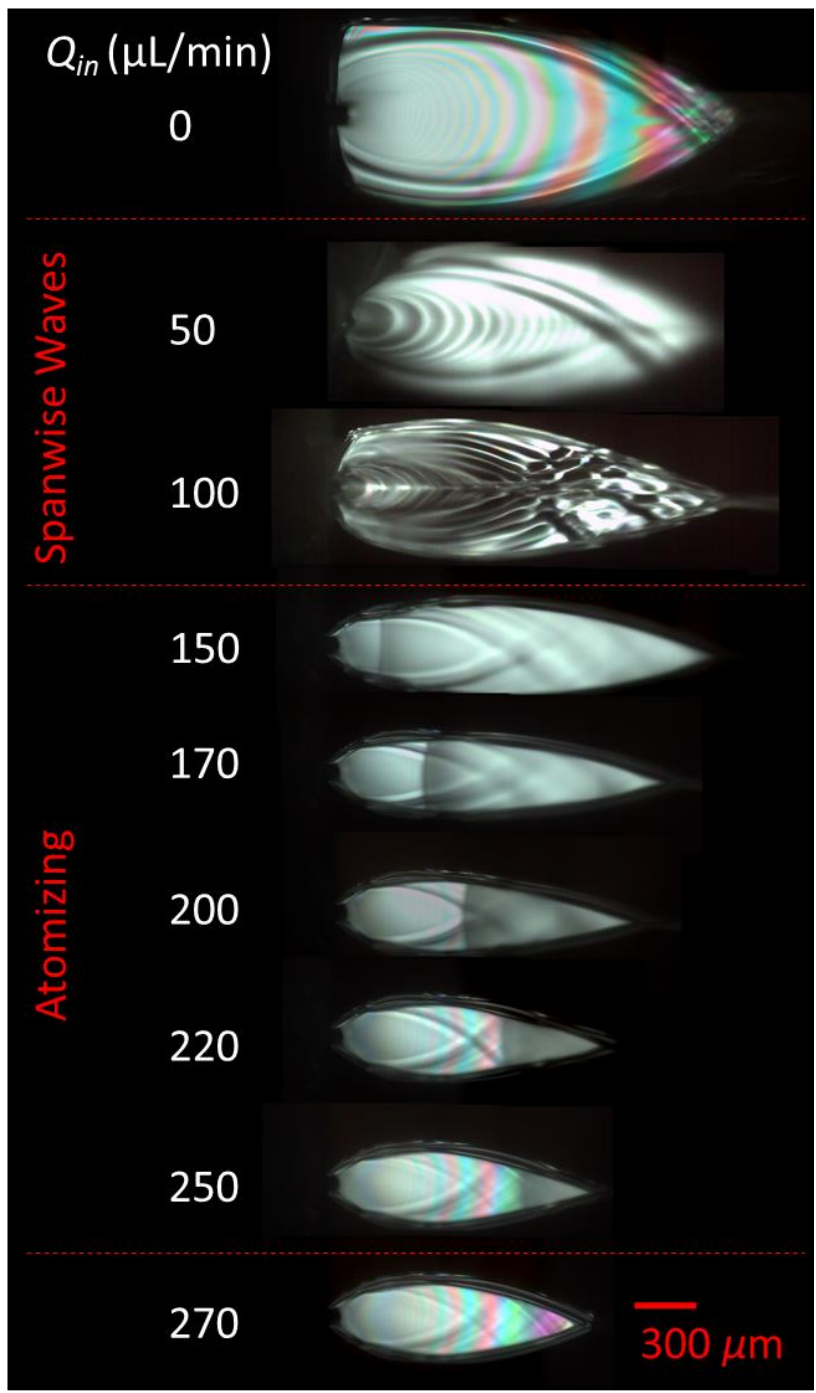

**Fig. S3.** Unstable to stable transition for Tol/Wat/Tol heterostructures ($Q_{out}$ = 1750 µL/min). The sheets in the "atomizing" region transition from stable to unstable midway through the sheet, possibly creating an emulsion within the sheet.

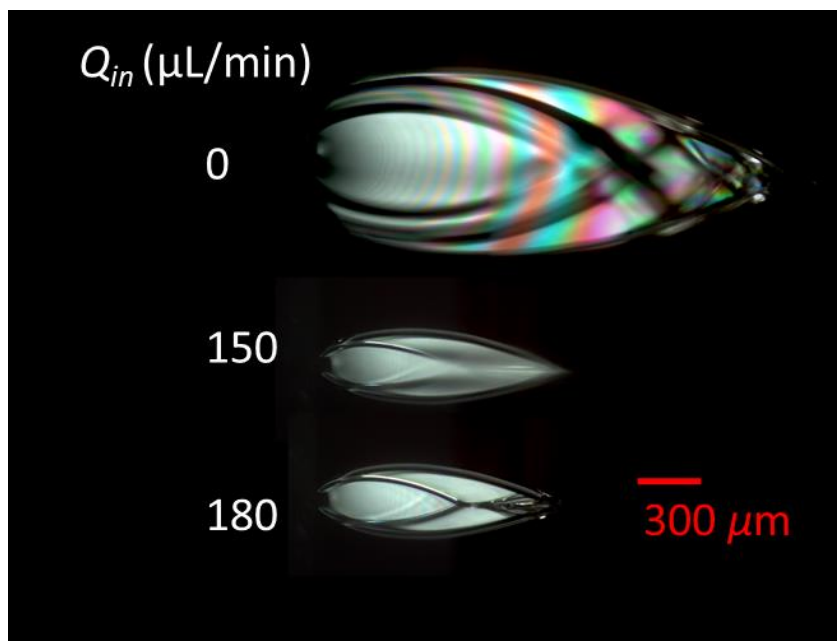

**Fig. S4.** Unstable to stable transition for Cy/Wat/Cy heterostructures ($Q_{out} = 1750\ \mu L/min$), showing the breakup of the small area sheet at low flow rates.

## S4. Additional FTIR Microscopy Images of Heterostructures

Representative false color images of various heterostructure morphologies are shown in Fig. S5. Red indicates the hydrocarbon component and blue indicates the water component. In general, it can be seen that all of the hydrocarbon/water/hydrocarbon heterostructures have the water rims contained within the sheet. Even in the low-$Q_{in}$ Tol/Wat/Tol structure where the layers fill the same area, the water rims are distinct from the toluene rims. By contrast, the water/hydrocarbon/water structures have rims which overlap in the image. The exact morphology of the water/hydrocarbon/water rims seem to be more complicated, and may relate again to interfacial tension. For the water/hydrocarbon/water structures, it is energetically favorable for the hydrocarbon to form an interface with air rather than water. By contrast, in the hydrocarbon/water/hydrocarbon structures, we expect that the water energetically favors forming an interface with the hydrocarbon instead of air. However, the exact structure of the

S.7

water/hydrocarbon/water rims remains ambiguous, and it is not clear that hydrocarbon/air interfaces are present at the rims of these structures.

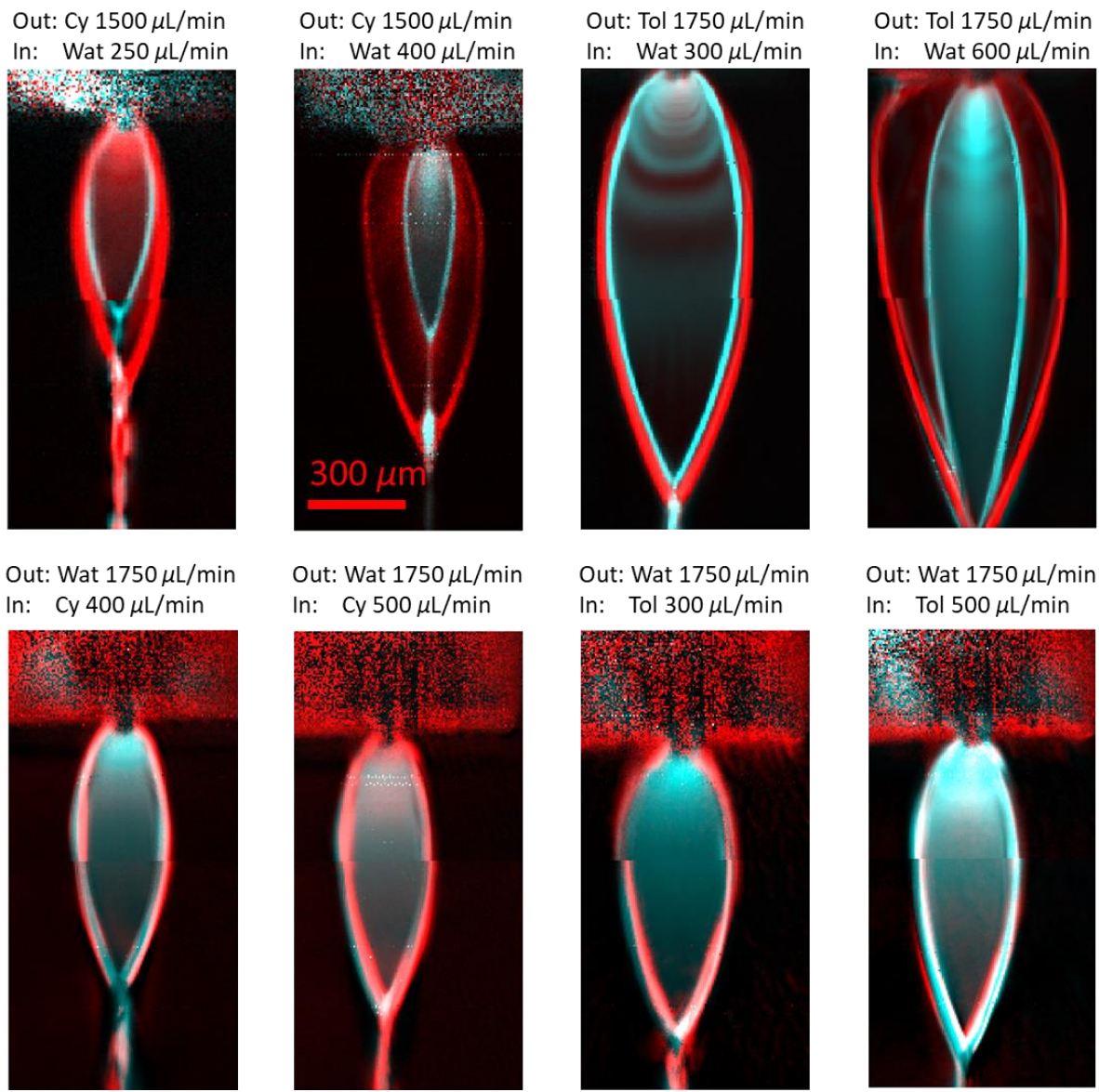

**Fig. S5.** Representative false color IR images of liquid heterostructures, with red indicating hydrocarbon and blue indicating water content. The water rims in hydrocarbon/water/hydrocarbon sheets are always within the sheet, while the hydrocarbon rims are coincident with the water rims in the water/hydrocarbon/water sheets.

## S5. Thin Film Calculations: Transfer Matrix Method

### A. Transfer Matrix Method

The transfer matrix method (TMM)[6, 7] was used to model the experimental data on the liquid heterostructures with IR microscopy, ellipsometry, and reflection measurements. TMM is well suited for describing the propagation of light through a stratified medium, which is the expected morphology for the liquid heterostructures. At each interface, the incident light can either transmit or reflect with coefficients given by the Fresnel equations depending on the incident polarization of light:

$$r_s = \frac{n_1 \cos\theta_1 - n_2 \cos\theta_2}{n_1 \cos\theta_1 + n_2 \cos\theta_2} \quad r_p = \frac{n_2 \cos\theta_1 - n_1 \cos\theta_2}{n_2 \cos\theta_1 + n_1 \cos\theta_2}$$
$$t_s = \frac{2n_1 \cos\theta_1}{n_1 \cos\theta_1 + n_2 \cos\theta_2} \quad t_p = \frac{2n_1 \cos\theta_1}{n_2 \cos\theta_1 + n_1 \cos\theta_2} \quad (S1)$$

where $n_m$ is the (complex-valued) index of refraction of the $m^{th}$ layer, and $\theta_m$ is the angle of the propagation vector relative to the interface normal. $\theta_m$ for each layer is derived from Snell's law:

$$n_1 \sin\theta_1 = n_2 \sin\theta_2. \quad (S2)$$

TMM uses the boundary conditions set by the interfaces with Maxwell's equations to efficiently solve the reflected and transmitted wave amplitudes. Full derivations can be found elsewhere in the literature.[6] The key result of TMM relates the forward and backward propagating wave amplitudes ($v$ and $w$, respectively) at the $m^{th}$ interface to the amplitudes at the $(m+1)^{th}$ interface by:

$$\begin{pmatrix} v_m \\ w_m \end{pmatrix} = M_m \begin{pmatrix} v_{m+1} \\ w_{m+1} \end{pmatrix} \quad (S3)$$

Where $M_m$ is a 2x2 matrix of the form:

$$M_m = \frac{1}{t_{m,m+1}} \begin{pmatrix} \exp(-i\delta_m) & 0 \\ 0 & \exp(+i\delta_m) \end{pmatrix} \begin{pmatrix} 1 & r_{m,m+1} \\ r_{m,m+1} & 1 \end{pmatrix}. \quad (S4)$$

The acquired phase difference $\delta_m = \frac{2\pi n_m \cos\theta_m}{\lambda_{vac}} d_m$ also takes into account the layer thickness $d_m$ and vacuum wavelength of light, $\lambda_{vac}$. This term also accounts for absorption by the medium when the index of refraction $n_m$ is complex-valued. The three layer, four interface liquid heterostructure reflection and transmission amplitudes can then be calculated from:

$$\begin{pmatrix} 1 \\ r \end{pmatrix} = \tilde{M} \begin{pmatrix} t \\ 0 \end{pmatrix}$$

$$\begin{pmatrix} 1 \\ r \end{pmatrix} = \frac{1}{t_{0,1}} \begin{pmatrix} 1 & r_{0,1} \\ r_{0,1} & 1 \end{pmatrix} M_1 M_2 M_3 \begin{pmatrix} t \\ 0 \end{pmatrix}. \tag{S5}$$

The matrix elements of $\tilde{M}$ then directly relate to the transmitted and reflected amplitudes of the full heterostructure, with $t = 1/\tilde{M}_{00}$ and $r = \tilde{M}_{10}/\tilde{M}_{00}$. The transmission and reflection can be calculated for either $s$ or $p$ polarization by using the appropriate Fresnel coefficients for the reflection and transmission.

## B. Application to FTIR Microscopy

As was noted in the main text, thin film interference effects produced a frequency-dependent baseline in the FTIR microscopy spectra. This effect was most prominent for Tol/Wat/Tol heterostructures, due to the weak absorption of toluene in the infrared region examined and the large area of the heterostructures compared to the Cy/Wat/Cy heterostructures. Correcting for the oscillatory baseline is also most important for structures with water as the interior liquid, as the strongest water mode (the OH bend at 3400 cm$^{-1}$)[1] spans several hundred wavenumbers, which makes it impractical to approximate the absorption baseline as locally flat.

The FTIR measurement was performed in transmission, where the transmitted intensity through the heterostructure for a given wavelength of light is:

$$T = |t|^2. \tag{S6}$$

(For this case, the transmitted intensity's dependence on index of refraction and angle of incidence drops out as the initial and final media are both air. As a result, there is no impedance mismatch between the final and initial medium). This intensity can be readily converted into an absorption value, A, by taking the logarithm:

$$A = -\log_{10}(T). \tag{S7}$$

The full absorption spectrum can be fully modeled by calculating $A$ for each wavelength using the complex-valued index of refraction for each liquid as a function of frequency. As the sheets were measured at a normal angle of incidence, the $s$ and $p$ Fresnel coefficients are identical, so polarization does not need to be considered.

The oscillatory background is only apparent if the light is coherent: as the layer thickness becomes large relative to the wavelength of light, the fringes become close enough together that they blend together due to minor thickness variations, propagation angle variations, etc. As the layers get thicker, the interference background transitions from an oscillatory coherent background to a flat incoherent background. This effect was accounted for by breaking the absorption spectrum into dispersive and absorptive parts. The coherent, oscillatory background, $A_{bg}$, was calculated using the TMM method using the real-valued portion of the index of refraction. The experimental absorption spectrum was then modeled as:

$$A = A_{inc} + fA_{bg} + A_{abs} \tag{S8}$$

Where $A_{inc}$ is the frequency-independent incoherent baseline, $A_{bg}$ is the frequency-dependent coherent baseline, $f$ is a scalar ranging from 0 to 1 to account for the coherence length, and $A_{abs}$ is the IR absorption spectrum for an incoherent film. This formulation technically does not account for absorption that occurs on internal reflections within the heterostructure, but the liquids used are only weakly reflecting at normal incidence. The resulting error from this simplification should be on the order of 1%.

A sample fit to an experimental spectrum is shown in Fig. S6 (Tol/Wat/Tol structure with $Q_{out} = 1750$ μL/min and $Q_{in} = 300$ μL/min, blue curve in Fig. 4A in main text). The red dashed curve is the full fit with Eq. S8, while the blue dashed curve is the baseline from $A_{inc} + fA_{bg}$. The blue dashed curve was subtracted to get the baseline corrected absorption spectra shown in Fig. 4A in the main text.

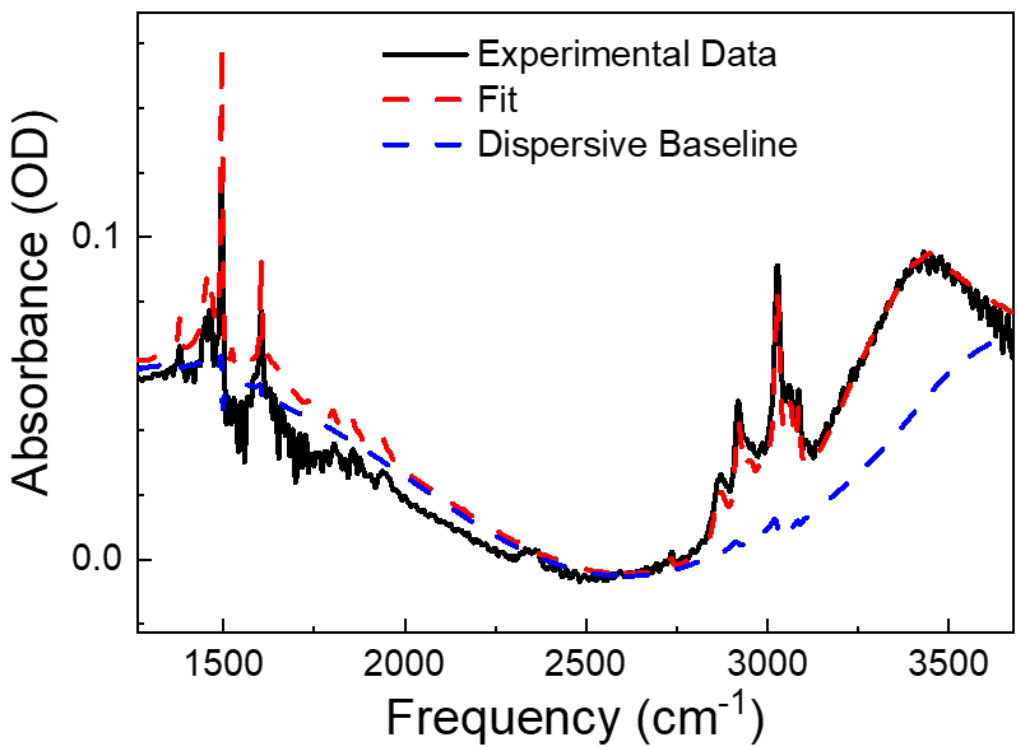

**Fig. S6.** Accounting for oscillatory IR background from thin film interference. Red dashed curve is the fit to the experimental spectrum with Eq. S8, while the blue curve indicates the dispersive baseline that was subtracted to generate the pure absorptive spectra in Fig. 4 of the main text.

## C. Application to Ellipsometry

The ellipsometric angles $\Delta$ and $\Psi$ are readily calculable using TMM. After $\rho = r_p/r_s$ are calculated for the heterostructure, the angles can be calculated from:

$$\Psi = \arctan(|\rho|)$$
$$\Delta = -i \log\left(\frac{\rho}{|\rho|}\right). \quad (S9)$$

For the liquid heterostructures, the values of both $\Delta$ and $\Psi$ have fine structure which arises in part from the dark bands: the points in the sheet where the four reflections fully destructively interfere. At these points, $\rho = 0/0$ and the ellipsometric angles are undefined. To construct the reference curves used to describe $\Psi$ in the main text, the values of $\Psi$ at the points of maximum constructive interference were used and interpolated between. The reflectivity and full calculated $\Psi$ value for the case shown in the main text (Tol/Wat/Tol structure with $Q_{out} = 1750\ \mu$L/min and $Q_{in} = 300\ \mu$L/min at a 70° AOI) is demonstrated in Fig. S7. The reference curves shown in Fig. 7B were generated by interpolating between the values of $\Psi$ at the points of maximum reflectivity in alternating bands. (Fig. S7, red and black curves). The experimental data was similarly analyzed at the points of maximum reflectivity, as that provided the highest signal to noise.

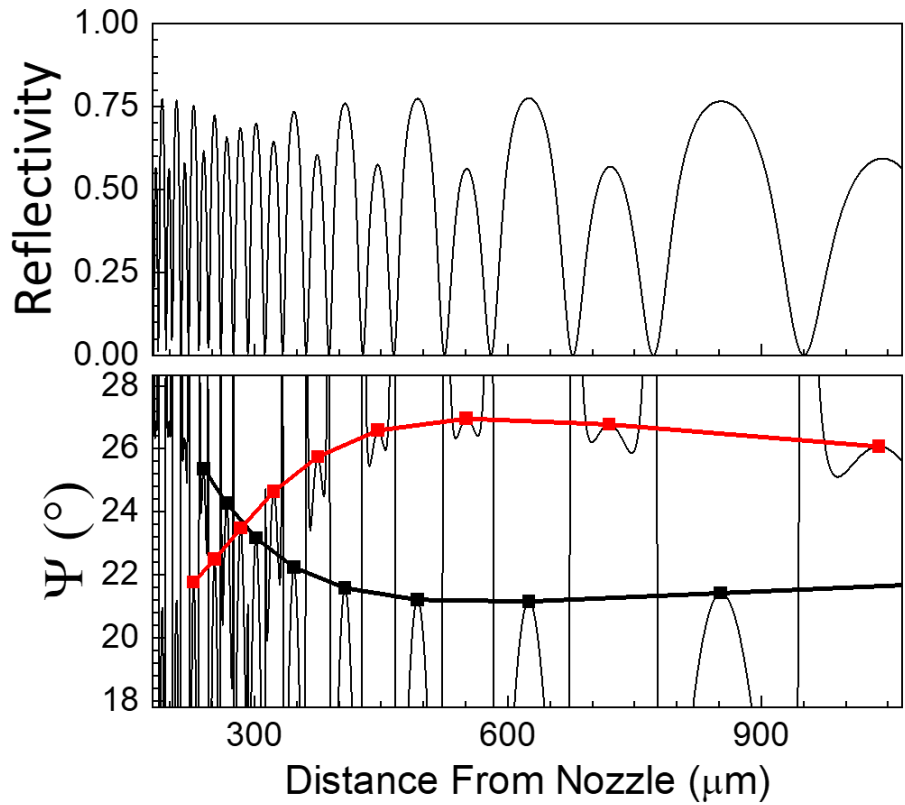

**Fig. S7.** Calculated reflectivity and Ψ values for the Tol/Wat/Tol structure with $Q_{out}$ = 1750 μL/min and $Q_{in}$ = 300 μL/min at a 70° AOI. The fit curves used in the main text was generated by interpolating between the values of Ψ at the maximum reflectivity for alternating bands.

### D. Application to White Light Reflectivity

To generate the white light color bar shown in Fig. 8, first the reflected intensity from the sheet was calculated for each visible wavelength:

$$I_{ref} = RI_{in}$$
$$I_{ref} = |r|^2 I_{in}$$
(S10)

The value of $I_{in}$ for each wavelength was weighted based on the spectrum of the white light LED used. As was the case for the FTIR background, the white light images were taken at near-normal incidence, so the polarization effects can be neglected. Each wavelength was then converted to an RGB value using a color-matching function (CIE 1964 10-degree CMF) at a 1 nm wavelength resolution. The CMF is a function which accepts a wavelength as input and returns coefficients for red, green, and blue channels. The calculated RGB values seen for a heterostructure with given layer thicknesses could then be determined using:

$$I_R = \frac{\sum_\lambda \text{CMF}_R(\lambda) I_{ref}(\lambda)}{\sum_\lambda I_{ref}(\lambda)} \tag{S11}$$

with corresponding expressions for the G and B channels.

An example full 2D color chart as a function of inner and outer layer thicknesses for a Wat/Tol/Wat heterostructure can be seen in Fig. S8A. In general, it can be seen that a very thin layer (< 100 nm) does not reflect any light, while a very thick layer (> 5 $\mu$m) reflects essentially white light, while layers around 1 $\mu$m thick give vibrant colors due to interference. Notably, a thick outer layer with a thin inner layer will produce a colored band against the white background corresponding to the thin inner layer. A thin inner layer is then noticeable in an otherwise thick part of a liquid heterostructure.

These characteristic colored bands of a thin inner layer could be generated in liquid heterostructures with low inner fluid flow rates. An example is shown in Fig. S8B for Wat/Tol/Wat liquid sheets. Near the top of the nozzle, the center line of the outer colliding sheet is at its thickest. In this case, the sheet is sufficiently thick that it would produce an essentially white reflection (e.g., bottom left of Fig. S8A). Introducing the inner liquid just above the threshold flow rate for generating a stable structure, the first colored band (dark yellow, bottom center of Fig. S8A) becomes visible near the top of the sheet (Fig. S8B left). As the inner flow rate increases, additional colored bands become visible near the top of the sheet, until at high flow rates the top of the sheet appears white again.

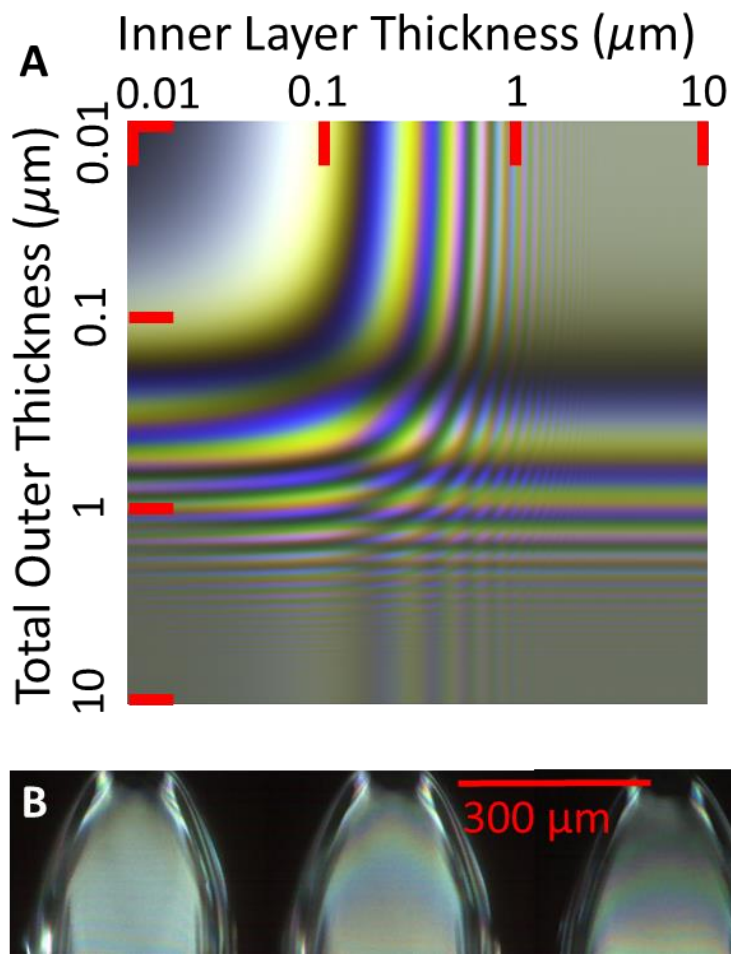

**Fig. S8. A**. Characteristic reflected color of a white LED off an ideal heterostructure as a function of inner and outer layer thickness. **B**. White light reflection off a Wat/Tol/Wat heterostructure with $Q_{out}$ = 1750 µL/min and a range of Qin. Thin film interference from a thin inner layer in a thick outer layer is visible near the top of the sheet (cf. bottom middle of B).

### S6. Polarization Modulation Imaging Ellipsometry Methods

#### A. Working Principle

The PM imaging ellipsometer was used based on the design by Chou et al.[8] The liquid sheet was illuminated with light polarized at 45°, meaning the incident light had equal *s* and *p* polarization components. The light reflected from the sheet is collected by an objective, after which it passes through a quarter-wave plate on a motorized stage and a linear polarizer set at

S.16

90° before reaching the CMOS camera. The field seen on each pixel of the camera can be described using Jones calculus:

$$\begin{bmatrix} E_{ref,p} \\ E_{ref,s} \end{bmatrix} = |E_{in}| \begin{bmatrix} 0 & 0 \\ 0 & 1 \end{bmatrix} \begin{bmatrix} \cos^2(\beta) + i\sin^2(\beta) & (1-i)\sin(\beta)\cos(\beta) \\ (1-i)\sin(\beta)\cos(\beta) & \sin^2(\beta) + i\cos^2(\beta) \end{bmatrix}$$
$$\times \begin{bmatrix} |r_p| & 0 \\ 0 & |r_s|e^{i\Delta} \end{bmatrix} \begin{bmatrix} 1 \\ 1 \end{bmatrix} e^{-i\pi/4}, \quad (S12)$$

where $\beta$ is the angle of the quarter wave plate's fast axis relative to the plane of incidence. The intensity seen by each pixel as a function of quarter-wave plate angle is then:

$$I_{ref}(\beta) \propto \frac{1}{4} \begin{bmatrix} 3|r_s|^2 + |r_p|^2 + \left(|r_s|^2 - |r_p|^2\right)\cos(4\beta) \\ -2|r_s||r_p|\left(2\sin(\Delta)\sin(2\beta) + \cos(\Delta)\sin(4\beta)\right) \end{bmatrix}. \quad (S13)$$

The ellipsometric angles $\Delta$ and $\tan\Psi = (|r_p|/|r_s|)$ can be determined using just five quarter-wave plate angles: $\beta = 0°$, 22.5°, 45°, 67.5°, and 135°:

$$\begin{aligned}
I_0 &= I_{ref}(0°) = |r_s|^2 \\
I_1 &= I_{ref}(22.5°) = \frac{1}{4}\left(3|r_s|^2 + |r_p|^2 - 2|r_s||r_p|\left(\sqrt{2}\sin\Delta + \cos\Delta\right)\right) \\
I_2 &= I_{ref}(45°) = \frac{1}{2}\left(|r_s|^2 + |r_p|^2 - 2|r_s||r_p|\sin\Delta\right) \\
I_3 &= I_{ref}(67.5°) = \frac{1}{4}\left(3|r_s|^2 + |r_p|^2 - 2|r_s||r_p|\left(\sqrt{2}\sin\Delta - \cos\Delta\right)\right) \\
I_4 &= I_{ref}(135°) = \frac{1}{2}\left(|r_s|^2 + |r_p|^2 + 2|r_s||r_p|\sin\Delta\right)
\end{aligned} \quad (S14)$$

which can be straightforwardly algebraically manipulated to get the desired quantities:

$$\begin{aligned}
\Psi &= \arctan\left(\sqrt{\frac{I_2 + I_4 - I_0}{I_0}}\right) \\
\Delta &= \arctan_2\left((I_4 - I_2), 2(I_1 - I_3)\right)
\end{aligned} \quad (S15)$$

where $\arctan_2$ is the two-argument arctangent function.

The results in Eqs. S14 and S15 differ from the results presented in Chou et al.[8] as the linear polarizer before the camera was set to 90° instead of 0°. This means the $r_s$ and $r_p$ terms switch places in Eq. S13 and S14, which then changes the expression in Eq. S15. This modification was made to reduce the dynamic range seen by the camera when the sheet was near Brewster's angle. In the Chou et al. configuration, the $\beta = 0°$ image is almost completely dark near Brewster's angle. In the configuration used here, the sheet illumination is more uniform between the collected quarter-wave plate angles.

**B. Image Processing**

Two significant artifacts were found in the collected images. First, dark regions did not read as identically zero, likely from a diffuse scattered light background and read noise on the camera. This background does not impact the calculation of $\Delta$, as it depends only on intensity differences, but it does impact $\Psi$. The image intensities were baselined to zero by selecting a region of the image which did not contain the sheet and subtracting the mean intensity of the region. Second, the rotation of the quarter-wave plate was not perfectly on-axis, resulting in a translation of the image by a few pixels between different $\beta$. This translation was corrected by observing a microscope grid distortion target while rotating the quarter-wave plate. The translation caused by the wave plate rotation was measured using the grid target. The measured translation which was then used to correct for the translation on the experimental data. These methods, as well as the set angles of the polarization optics and rotation stages, were validated using a step-wafer standard of known thickness and refractive index.

For displaying the ellipsometric data maps (e.g., in Figs. 6 and 7 in the main text), the sheet was highlighted relative to the dark background by using a mask. The mask was generated by weighing the sum of all five images using a sigmoid function, with parameters chosen so that the flat part of the sheet was uniformly white and the background was uniformly dark.

## Supplemental References

1. Venyaminov, S. Y.; Prendergast, F. G., Water (H2O and D2O) Molar Absorptivity in the 1000–4000 cm−1Range and Quantitative Infrared Spectroscopy of Aqueous Solutions. *Anal. Biochem.* **1997,** *248*, 234-245.

2. Lange, N. A.; Dean, J. A., *Lange's Handbook of Chemistry*. McGraw-Hill: 1973.

3. Backes, H. M.; Jing Jun, M.; E., B.; G., M., Interfacial tensions in binary and ternary liquid—liquid systems. *Chem. Eng. Sci.* **1990,** *45*, 275-286.

4. Ryan, H. M.; Anderson, W. E.; Pal, S.; Santoro, R. J., Atomization characteristics of impinging liquid jets. *J. Propul. Power* **1995,** *11*, 135-145.

5. Koralek, J. D.; Kim, J. B.; Brůža, P.; Curry, C. B.; Chen, Z.; Bechtel, H. A.; Cordones, A. A.; Sperling, P.; Toleikis, S.; Kern, J. F.; Moeller, S. P.; Glenzer, S. H.; DePonte, D. P., Generation and characterization of ultrathin free-flowing liquid sheets. *Nat. Commun.* **2018,** *9*, 1353.

6. Harbecke, B., Coherent and incoherent reflection and transmission of multilayer structures. *Appl. Phys. B* **1986,** *39*, 165-170.

7. Byrnes, S. J., Multilayer optical calculations. *arXiv preprint* **2016,** *arXiv:1603.02720* 1-20.

8. Chou, C.; Teng, H.-K.; Yu, C.-J.; Huang, H.-S., Polarization modulation imaging ellipsometry for thin film thickness measurement. *Opt. Commun.* **2007,** *273*, 74-83.